\documentclass[floats,tighten,amsmath,amssymb,twocolumn,prb]{revtex4}
\usepackage{graphicx}    
\usepackage{epsfig}      
\usepackage{dcolumn}     
\usepackage{bm}          
\usepackage{color}
\graphicspath{{./}{Figures//}}
\def\lmo{LaMnO$_3$}
\def\cmo{CaMnO$_3$}
\def\lcmo{La$_{1-x}$Ca$_{x}$MnO$_{3}$}
\def\mo6{MnO$_6$}
\def\rhm3{${R\overline 3c}$}
\def\pnma{$Pnma$}

\begin{document}
%
%
\begin{abstract}
Evolution of the average and local crystal structure of Ca-doped \lmo\ has been studied across the metal to insulator (MI) and the orthorhombic to rhombohedral (OR)
structural phase transitions over a broad temperature range for two Ca concentrations ($x=0.18, 0.22$). Combined Rietveld and high real space resolution
atomic pair distribution function (PDF) analysis of neutron total scattering data was carried out with aims of exploring the possibility of nanoscale phase
separation (PS) in relation to MI transition, and charting the evolution of local Jahn-Teller (JT) distortion of \mo6\ octahedra across the OR transition at T$_S$$\sim$720~K.
The study utilized explicit two-phase PDF structural modeling, revealing that away from T$_{MI}$ there is no evidence for nanoscale phase coexistence. The local JT-distortions
disappear abruptly upon crossing into the metallic regime both with doping and temperature, with only small temperature-independent signature of quenched disorder being
observable at low temperature as compared to \cmo. The results hence do not support the percolative scenario for the MI transition in \lcmo\ based on PS, and question its ubiquity in the manganites. In contrast to \lmo\ that exhibits long range orbital correlations and sizeable octahedral distortions at low temperature, the doped samples with compositions straddling the MI boundary exhibit correlations (in the insulating regime) limited to only $\sim$1~nm with observably smaller distortions. In $x=0.22$ sample local JT-distortions are found to persist across the OR transition and deep into the R-phase (up to $\sim$1050K) where they are crystallographically prohibited. Their magnitude and subnanometer spatial extent remain unchanged.
\end{abstract}
\title{Non percolative nature of the metal-insulator transition and persistence of local Jahn-Teller distortions in the rhombohedral regime of La$_{1-x}$Ca$_{x}$MnO$_{3}$}
\author{Mouath Shatnawi}
\affiliation{Department of Physics, The Hashemite University,
  Zarqa 13115, Jordan }
\author{Emil S. Bozin}
\affiliation{Department of Condensed Matter Physics and
  Materials Science, Brookhaven National Laboratory, Upton, New York
  11973, USA}
\author{J. F. Mitchell}
\affiliation{Materials Science Division, Argonne National Laboratory, Argonne, Illinois 60439 }
\author{Simon J. L. Billinge}
\affiliation{Department of Condensed Matter Physics and
  Materials Science, Brookhaven National Laboratory, Upton, New York
  11973, USA}
\affiliation{Department of Applied Physics and Applied
  Mathematics, Columbia University, New York, New York 10027, USA}
\date{\today}


\maketitle

%
\section{Introduction}
\label{intro}
%
Manganites, such as the perovskite structured calcium doped lanthanum manganite,
La$_{1-x}$Ca$_{x}$MnO$_{3}$ (LCMO), are
considered as a model system for studying the response of materials to
the presence of competing interactions,~\cite{milli;n98,yuno;prl98,zhou;prb03a,dagot;s05,beben;pmm11} yet despite
extensive study the emergent properties, such as the colossal magnetoresistance and the metal-insulator (MI) transition, in these materials are far from well understood.

Much recent work has focused on an inhomogeneous, or what is
called phase separation (PS),
picture~\cite{uehar;n99,dagot;ssc03,burgy;prl04,dagot;jpcm08,zhang;physs12,kuber;ass12,wu;pb12,phong;jec13}
where metallic domains are assumed to form in an insulating
background and the MI transition proceeds via a
percolative mechanism where the metallic cluster size increases as
the MI transition is approached resulting in a conducting path
extending from one side of the sample to the other. Although the PS
picture has been supported by many experimental studies such as
scanning tunneling spectroscopy,~\cite{fath;s99} small angle
neutron scattering,~\cite{teres;n97} atomic pair distribution
function (PDF)~\cite{billi;prb00} and nuclear magnetic
resonance,~\cite{allod;prb97} there are some experimental results
that are better explained based on the homogeneous
picture~\cite{kumar;prx14,mitra;prb05,tyson;prb96,hundl;apl95}
and the issue needs a more careful assessment.

\begin{figure}[tbp]
  \centering
 \includegraphics[angle=0, width=0.475 \textwidth]{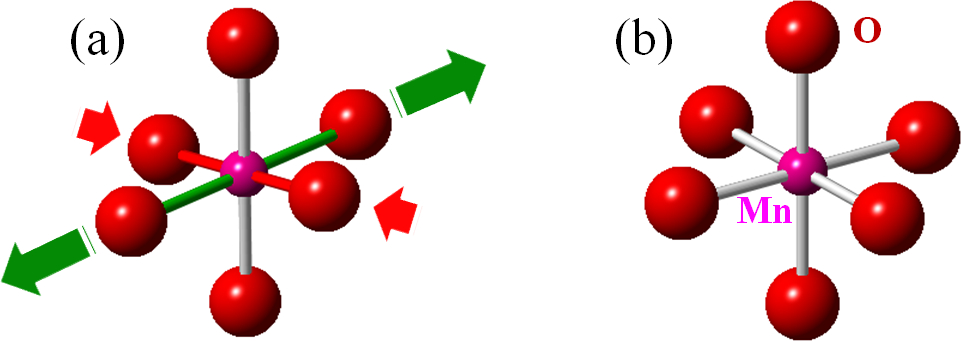}
\caption{(Color online) \mo6\ octahedra $-$ basic building block of the \lcmo\ structure: (a) JT-distorted, with pairs of long (green), intermediate
(gray), and short (red) Mn$-$O
distances; (b) undistorted, with all Mn$-$O distances equal (gray).}
  \label{fig;octahedron}
\end{figure}

Through the strong electron-lattice interaction, the presence or absence of a JT distortion has been shown to be a sensitive indicator of the
electronic state of the material; insulating or metallic~\cite{billi;prl96}.  It manifests itself in the Mn–O bond length ($r_{Mn-O}$)
distribution in the MnO$_{6}$ octahedron, irrespective of whether the distortions are long range ordered
or not~\cite{qiu;prl05,sartb;prl07}. When the JT distortion is present, the distribution of near neighbor
$r_{Mn-O}$ distances constituting the MnO$_6$ octahedron
will consist of two long ($\sim 2.16$~\AA\ in \lmo\ endmember), two intermediate and two short ($<2$~\AA)
Mn$-$O bonds (Fig.~\ref{fig;octahedron}(a)). When the JT distortion is absent there are six equal $r_{Mn-O}$
distances on the octahedron (Fig.~\ref{fig;octahedron}(b)).  Observation of the JT distortion in probes of local
structure are associated with the insulating phase~\cite{billi;prl96,kiryu;njp04,bozin;prl07}, whereas deep in the metallic regime
at base temperature no appreciable signature of the JT distortions have been observed on any length-scale~\cite{bozin;prl07}.

Here we apply the PDF method to study structural
changes in \lcmo\ as a function of temperature for two carefully
chosen concentrations: $x=0.18$ and $x=0.22$. These two
concentrations straddle the doping-induced MI transition, enabling
us to cross the MI transition both as a function of doping and
temperature. For $x=0.18$ the system exhibits insulating behavior
at all temperatures. For $x=0.22$, at low temperatures the system
is ferromagnetic metallic (FM). By heating to temperatures above
$\sim$180~K it becomes a paramagnetic insulator (PI).
The key result is that local JT distortions disappear abruptly
at the MI transition both as a function of doping at fixed temperature,
and by temperature at fixed doping. Explicit 2-phase modeling
reveals no evidence of phase coexistence away from T$_{MI}$.
This is incompatible with a percolative picture of the phase
transition.

Additionally, we have studied the transition to the high temperature
rhombohedral phase on heating in the $x=0.22$ material.  Local JT-distortions
persist at high temperature, although
they are crystallographically prohibited in the \rhm3\ phase,
similar to the earlier observation in the \lmo\
endmember~\cite{chatt;prb03,qiu;prl05}. Transition to the rhombohedral regime appears rather continuous
from the local structure perspective, with no observable change in
the size and spatial extent of the local distortions.

%
\section{EXPERIMENT AND MODELING}
\label{exp}
%
Finely pulverized samples of \lcmo\ with two compositions, $x=0.18$ and $x=0.22$, were prepared using standard solid state
synthesis methods~\cite{qiu;prl05}, annealed to ensure oxygen stoichiometry~\cite{dabro;jssc89}, and characterized
thoroughly~\cite{bozin;prl07}. Neutron powder diffraction measurements were carried out at the NPDF diffractometer
at Los Alamos Neutron Scattering Center at Los Alamos National Laboratory.
The samples, approximately 6 grams each, were loaded into extruded vanadium containers and sealed under He atmosphere.
The data were collected on warming for both samples at various temperatures between 10~K and 550~K using closed cycle cryo-furnace.
In addition, in order to explore the atomic structure at various length-scales within the high temperature rhombohedral phase,
data on $x=0.22$ sample were also collected on warming in ILL furnace up to a temperature of 1050~K. Transformation to rhombohedral phase
for this composition occurs at $\sim$ 720~K~\cite{salam;rmp01}. Control furnace measurement at 550~K was conducted on cooling to ensure the sample integrity.
Data were collected for 4 hours at each temperature on each sample, providing good statistics and a favorable signal to noise ratio at high momentum transfers.
Raw data were normalized and various experimental corrections performed following standard reduction protocols~\cite{egami;b;utbp12}.
High resolution experimental PDFs were obtained from the Sine Fourier transform of the measured total scattering structure functions, $F(Q)$, over a broad range
of momentum transfers up to $Q_{\rm max}  = 35$~\AA$^{-1}$. Data reduction to obtain the PDFs, $G(r)$, was carried out using the program {\sc PDFGETN}~\cite{peter;jac00}.

The average structure was assessed through the Rietveld refinements~\cite{rietv;ac67} to the raw
diffraction data using {\sc GSAS}~\cite{larso;unpub87} operated under {\sc EXPGUI}~\cite{toby;jac01}.
Structural refinements of the PDF data were carried out over $1.7$~\AA\ $< r < 3.5$~\AA\ range using {\sc PDFfit2} operated under {\sc PDFgui}.\cite{farro;jpcm07}
In addition, for PDF data at selected temperatures refinements were also carried out using a variable $r_{max}$ protocol~\cite{qiu;prl05}.
In this, the low-$r$ refinement bound was kept fixed at 1.7~\AA, while the upper-$r$ bound was systematically changed from 3.5~\AA\ to 20~\AA\ to
provide an estimate of the correlation length of the locally ordered distortions.

The average structure refinements were carried out by the known structural models appropriate for the temperature and concentration ranges used.
The $O'$ structural model (\pnma) was used for temperatures up to 650~K~\cite{bozin;jpcs08} and the $R$ structural model (\rhm3) was used
for the 750~K$-$1050~K range.~\cite{huang;prb97,souza;prb08} The local and intermediate structure refinements of PDF data for all temperatures
were carried out with $O'$ structural model. All refinements were carried out with isotropic atomic displacement parameters (ADP).
Two-phase refinements were carried out to explore the phase separation scenario: both phases were described within the $O'$ structure,
with structural parameters fixed to values obtained deep in the metallic and insulating regimes, and only parameters allowed to vary were the
phase fraction and the atomic ADPs. The later were constrained to be the same for related atomic sites across the two phases.

\section{RESULTS AND DISCUSSION}
\label{results}
%
\begin{figure}[tbp]
  \centering
 \includegraphics[angle=0, width=0.475 \textwidth]{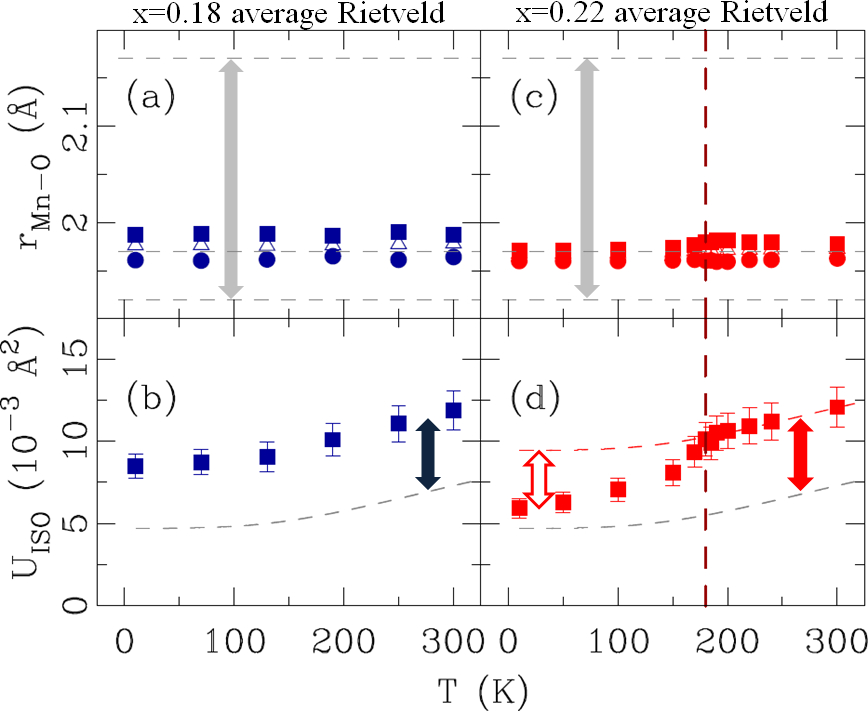}
\caption{(Color online) Average structure behavior (Rietveld). Top row: Temperature evolution of the average Mn$-$O distances of \mo6\ octahedron for (a) $x=0.18$ (blue symbols) and
(c) $x=0.22$ (red symbols). Horizontal dashed lines represent these distances in \lmo\ end-member with vertical gray double-arrow indicating the distortion size. Bottom row:
Temperature evolution of uncorrelated isotropic atomic displacement parameter (ADP) of 8d oxygen (\pnma\ model) for (b) $x=0.18$ (blue symbols) and (d) $x=0.22$ (red symbols). Sloping
dashed gray lines denote uncorrelated isotropic ADPs of the same oxygen site in \cmo\ end-member as a reference in the absence of disorder.~\cite{bozin;jpcs08} Sloping dashed red line
in (d) represents the same, but with added constant offset to match the high temperature end of the ADP data. Double arrows indicate excess disorder in doped samples as compared to
\cmo. Vertical dashed red line marks T$_{MI}$.
}
  \label{fig;RietveldDistancesADPs}
\end{figure}
%
\subsection{Structural evolution across the MI transition from single phase modeling}
\label{MIT}
%
%
\subsubsection{Average and local structure}
\label{aveInterStruct}
%
First we briefly review the behavior of Mn-O bond lengths and isotropic ADPs of oxygen
when modeled using a single phase model that was previously described
and reported in earlier work~\cite{bozin;prl07}.
\begin{figure}[tbp]
  \centering
\includegraphics[angle=0, width=0.45 \textwidth]{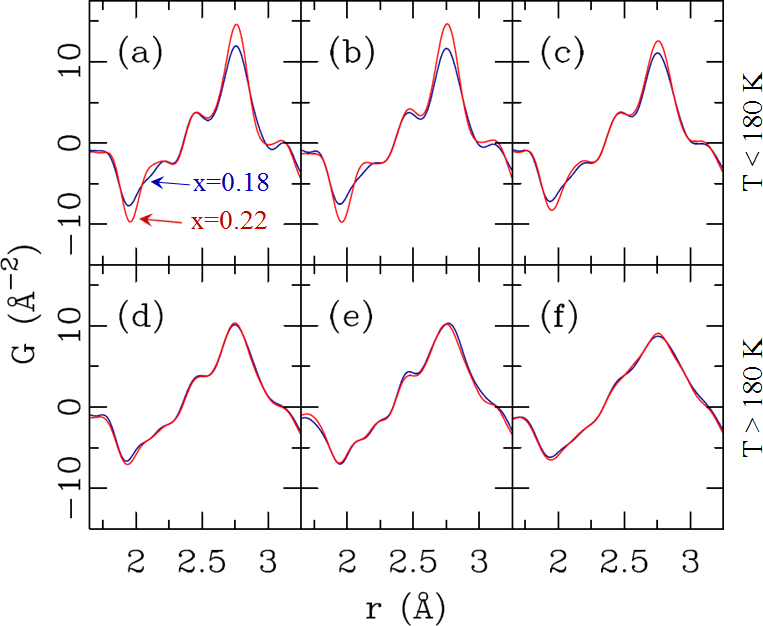}
\caption{(Color online) Comparison of experimental PDFs over r-range describing \mo6\ octahedron, for $x=0.18$ (solid blue line) and $x=0.22$ (solid red line) at various comparable temperatures: T $<$ T$_{MI}$ (top row), and T $>$ T$_{MI}$ (bottom row). (a) 10~K, (b) $\sim$50~K, (c) $\sim$150~K, (d) $\sim$250~K (e) 300~K, and (f) 550~K.
Notable difference can be observed at low temperature where $x=0.18$ sample is insulating, whereas $x=0.22$ sample is metallic. At T $>$ T$_{MI}$ visible difference vanishes.
}
  \label{fig;PDF18vs22}
\end{figure}

Figure~\ref{fig;RietveldDistancesADPs} summarizes the situation from the average structure perspective (Rietveld refinement). At
these compositions in 10~K$-$ 300~K temperature range the samples show a very small, or negligible,~\cite{salam;rmp01,chatt;prb02,kiryu;prb03,kiryu;prb04,kiryu;njp04}
JT distortion with all 6
Mn$-$O bonds in the octrahedra having a similar length in the range around 1.975~\AA .  For reference, the Mn$-$O
bonds in the undoped endmember, where the JT distorted octahedra are long-range ordered, are 2.16~\AA , 1.96~\AA , and 1.93~\AA ,
as shown in the figure as the dashed lines.  On the other hand, an enlarged ADP of oxygen on the octahedra indicates that there is some
disorder associated with the oxygen positions, consistent with the presence of disordered JT distorted octahedra (for comparison
the isotropic ADP of the same oxygen site when no structural disorder is present, in the \cmo\ end-member~\cite{bozin;jpcs08}, is shown
as the dashed line)~\cite{radae;prb96i}.
The value of the oxygen ADP rapidly decreases in the $x=0.22$ sample as it enters the metallic phase at low temperature,
approaching, but not reaching, the undistorted value (Fig.~\ref{fig;RietveldDistancesADPs}(d)).
The slightly enhanced value of this ADP at base temperature  is presumably a signature
of the quenched disorder component due to Ca/La substitution, which is a temperature independent quantity.
Refinements of the PDF over a range $1.7<r<40$~\AA\ (not shown) quantitatively agrees with the Rietveld results.

\begin{figure}[tbp]
  \centering
\includegraphics[angle=0, width=0.45 \textwidth]{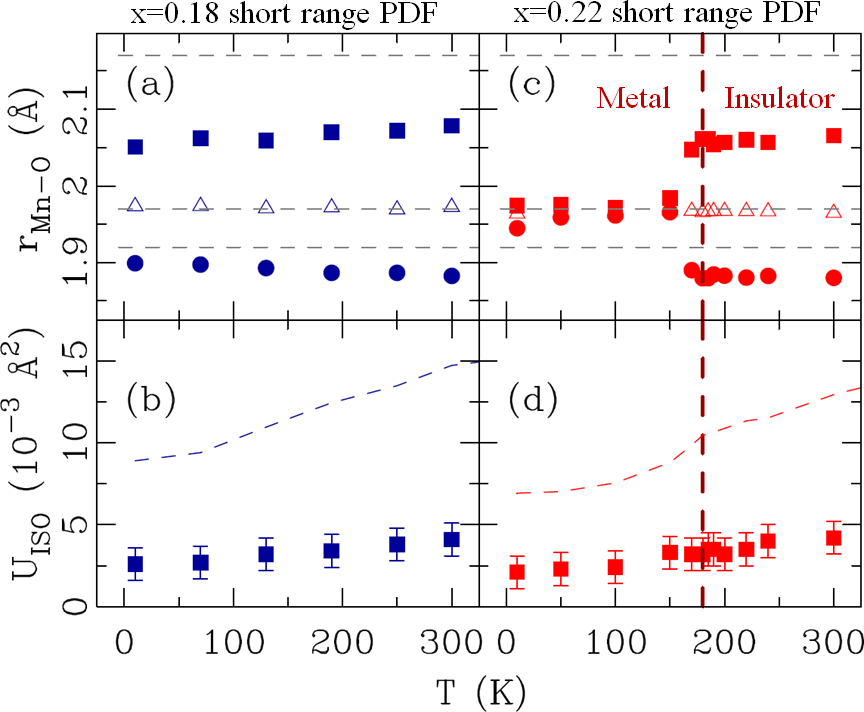}
\caption{(Color online) Local structure behavior (PDF). Top row: Temperature evolution of the local Mn$-$O distances of \mo6\ octahedron for (a) $x=0.18$ (blue symbols) and (c)
$x=0.22$ (red symbols). Horizontal dashed lines represent these distances in \lmo\ end-member. Bottom row: Temperature evolution of correlated isotropic ADPs of 8d oxygen (\pnma\
model) for (b) $x=0.18$ (blue symbols) and (d) $x=0.22$ (red symbols). Dashed lines sketch the uncorrelated ADPs shown in Fig.~\ref{fig;RietveldDistancesADPs}(b) and (d) for reference.
Vertical dashed red line marks T$_{MI}$.
}
  \label{fig;PDFDistancesADPs}
\end{figure}

Examination of the measured PDFs in the low-$r$ region of the data confirm that the enlarged ADPs seen in Rietveld are coming from the presence of
non-long-range-ordered JT distorted octahedra.
This is most effectively demonstrated by comparison between the experimental PDFs when the PDF profiles
of the two concentrations are plotted on top of each other, as shown in Fig.~\ref{fig;PDF18vs22}. For temperatures lower than
$T_{MI}$ (Figs.~\ref{fig;PDF18vs22}(a), (b) and (c)) the PDFs for the two samples show different features. For higher
temperatures (Figs.~\ref{fig;PDF18vs22}(d), (e) and (f)) the PDFs of $x=0.18$ and $x=0.22$ are very similar.
For both samples and at all measured temperatures the PDF peak corresponding to the Mn$-$O bonds on the MnO$_6$ octahedra
can bee seen around 2~\AA . This appears upside down, as a negative peak, due to the negative scattering length of
Mn~\cite{egami;b;utbp12}. Further, the compound peak at around 2.75 ~\AA\ includes the O$-$O distances on the octahedron~\cite{bozin;jpcs08}.
The first PDF peak for $x=0.18$ and for all measured temperatures exhibits a shoulder on the high $r$ side of the
peak. Since at low temperatures the thermal broadening effect will be minor, the appearance of this shoulder
indicates the existence of a longer Mn$-$O bond, and therefore the presence of a JT distortion.
For $x=0.22$ below 180~K (MI transition temperature) the Mn$-$O peak appears as a single, well defined,
peak. Additionally, the compound peak describing O$-$O distances on the \mo6\ octahedron is sharp for $x=0.22$ (undistorted \mo6)
and broad for $x=0.18$ (distorted \mo6).
These structural features may be used to monitor changes in the JT content as the sample passes through the MI transition.
As the measurement temperature of the $x=0.22$ sample is increased
above 180~K a shoulder starts to appear on the high $r$ side similar to the one seen in $x=0.18$ sample coinciding
with the MI transition observed in transport~\cite{bozin;prl07}.
This was previously pointed out in the context of a temperature dependent study
in \lcmo\,~\cite{billi;prl96} but we show that the effect is similarly seen when
the MI transition is crossed as a function of doping at constant temperature.

Quantitative analysis from PDF refinements carried out over a narrow $r$-range,
reveal the distribution of bond-lengths within the local JT distorted octahedra
(Fig.~\ref{fig;PDFDistancesADPs}(a) and (c)), with
 ADPs of 8d-O associated with these fits shown in Fig.~\ref{fig;PDFDistancesADPs}(b) and (d).
The latter show no anomalies at the MI transition temperature since
all the information about the local distortions has been captured by
the structure parameters rather than the ADPs~\cite{bozin;prl07}.

These observations reinforce the well known result that the MI transition is accompanied by the formation
of local JT-distortions in the insulating phase, and, conversely, undistorted octahedra in the metallic regime.
It also demonstrates that the local bond-length distribution in the insulating regime
decreases in amplitude with increasing Ca content, as opposed to the assumption of the small polaron
picture in which the local octahedral distortion is as large as in the \lmo\
endmember~\cite{booth;prl98,louca;prb97,billi;prb00} associated with the charge
localized Mn$^{3+}$ sites, while Mn$^{4+}$ sites are undistorted as in \cmo.

%
\subsubsection{Correlation length of local JT-distortions}
\label{correlationLength}
%
\begin{figure}[tbp]
  \centering
 \includegraphics[angle=0, width=0.4 \textwidth]{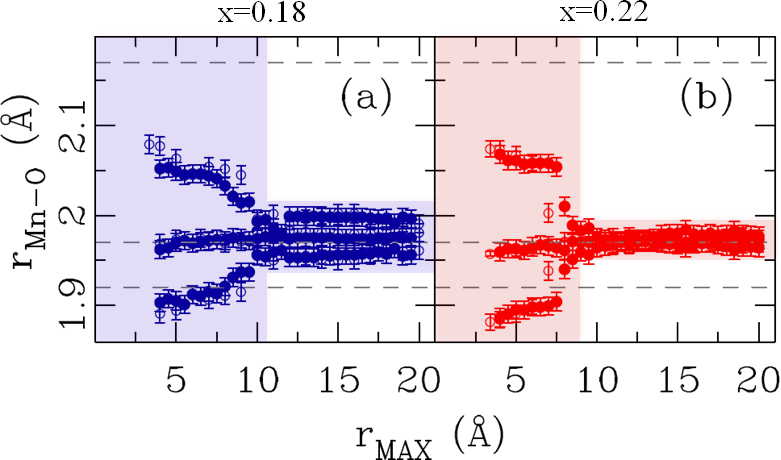}
\caption{(Color online) Mn$-$O bond lengths from the refined structure model to PDF data as a function of upper refinement limit, r$_{max}$, for (a) $x=0.18$ and (b) $x=0.22$
compositions. Assessment for $x=0.18$ was done at 10~K (solid symbols) and 300~K (open symbols), whereas for $x=0.22$ at 300~K for PDF data obtained at Q$_{max}$ values of
25~\AA$^{-1}$ (open symbols) and 35~\AA$^{-1}$ (solid symbols). Dashed lines indicate Mn$-$O bondlengths in \lmo\ at 300~K, where octahedral distortions are long range ordered. See
text for details.
}
  \label{fig;corrLength}
\end{figure}
The PDF yields the structure on different length-scales, revealing the locally JT distorted octahedra at
low-$r$ and the average, undistorted, structure at high-$r$.  By studying how the PDF signal crosses over
from the distorted to undistorted as a function of $r$ we may extract a correlation length for
any local ordering of the JT distorted octahedra.
A variable range PDF refinement of the 300~K data was performed from $r_{min}=1.7$~\AA\ to $r_{max}$, where $r_{max}$ was increased from
3.5~\AA\ to 20~\AA\ in steps of 0.5~\AA. The $r_{Mn-O}$ bond-lengths extracted from these refinements
are shown in Fig.~\ref{fig;corrLength}.
It can be seen that the amplitude of the measured octahedral distortion falls off gradually with
increased fitting range until the $r_{Mn-O}$ distances reach the values from the crystallographic
analysis.  The correlation length of the local JT distortions is $\sim 10.5$~\AA\ at $x=0.18$, decreasing to $\sim 8$~\AA\
at $x=0.22$.
This is very similar to the observation above the JT-transition in \lmo~\cite{qiu;prl05},
albeit the length-scale in the present case appears to be smaller, suggesting the decrease
of the local distortion correlation length with Ca-doping.

We have tested whether the length-scale
of ordered local distortions changes with temperature for the $x=0.18$ sample. Fig.~\ref{fig;corrLength}(a) shows
results of the assessment at 10~K and 300~K and we could not detect a temperature evolution of the characteristic length-scale
within the accuracy of the approach. We also tested whether the observations are dependent on the $r$-resolution of the PDF
for $x=0.22$ by doing the analysis from PDFs obtained using Q$_{max}$ values of 25~\AA$^{-1}$ and 35~\AA$^{-1}$ (open and solid symbols, respectively, in
Fig.~\ref{fig;corrLength}(b)). We find that slightly sharper and larger length-scale (by $\sim$ 1~\AA) is obtained with higher resolution
data, giving us an estimate of the uncertainty on this value. While the absolute accuracy of the applied protocol
could be somewhat challenging~\cite{bozin;sr14},
it still provides very important insights about the average correlation length of the underlying ordered local
distortions.
%
\subsection{Phase separation}
\label{phaseSeparation}
%
\begin{figure}[tbp]
  \centering
\includegraphics[angle=0, width=0.45 \textwidth]{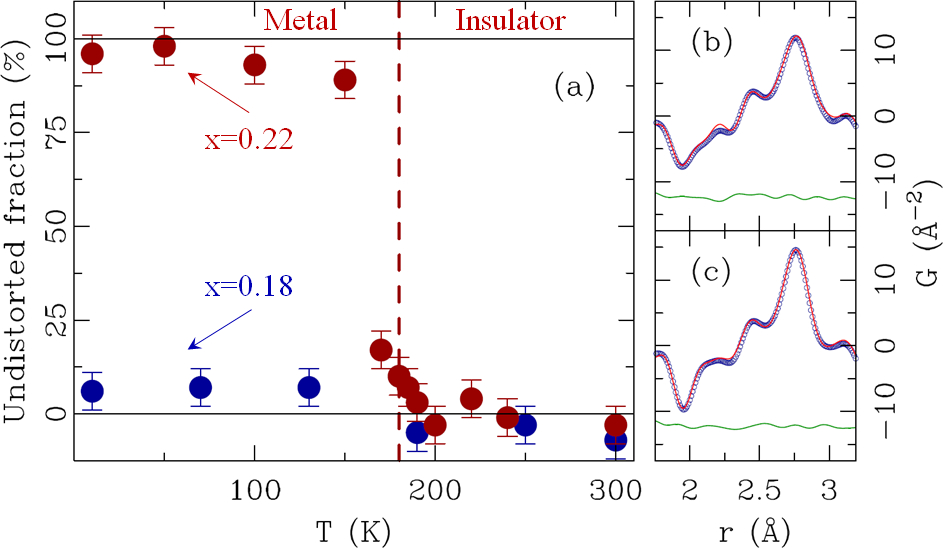}
\caption{(Color online) (a) T-evolution of undistorted fraction from 2-phase short-range PDF modeling for $x=0.18$ (solid blue symbols, no MIT), and for $x=0.22$ (solid red symbols, MIT at $\sim$ 180~K indicated by vertical dashed red line). Horizontal solid black lines are guides to the eye. Typical short range refinements
of the 2-phase model are shown in (b) for $x=0.18$ and in (c) for $x=0.22$ data at 10~K. In both panels open blue symbols represent the data, solid red line is the model, and
solid green line is the difference (offset for clarity).
}
  \label{fig;fraction}
\end{figure}

We now turn to the question of phase separation and  what our data have to say on this topic.
Whilst phase separation with large length-scale domains is clearly dominating in systems with smaller A-site ions such as La$_{1-x}$Pr$_x$MnO$_3$~\cite{marti;prb99}, for the systems that show robust CMR behavior such as Ca and Sr doped \lmo , though the phase separation scenario has been invoked, our results suggest that phase coexistence is minimal in the vicinity of the MI transition at $x\sim 0.2$, at least for the Ca doped state.
The rapid crossover from distorted to completely undistorted as the MI transition is crossed, both as a function of $T$ at $x=0.22$ and as a function
of doping, $x$ at fixed temperature, would seem to rule out a percolation mechanism for that transition.

Observations of phase separation in these systems from other measurements may be a result of extrinsic effects such as doping or strain, or difficulties
in data interpretion~\cite{uehar;n99,fath;s99,louca;prb97,becke;prl02,billi;prb00,rames;ml09,ward;prb11,tao;pnas11}.

The situation may be different at higher doping, such as La$_{0.5}$Ca$_{0.5}$MnO$_{3}$~\cite{loudo;n02}, and our
results only explicitly address the low-doped region of the phase diagram.

\begin{figure}[tbp]
  \centering
\includegraphics[angle=0, width=0.475 \textwidth]{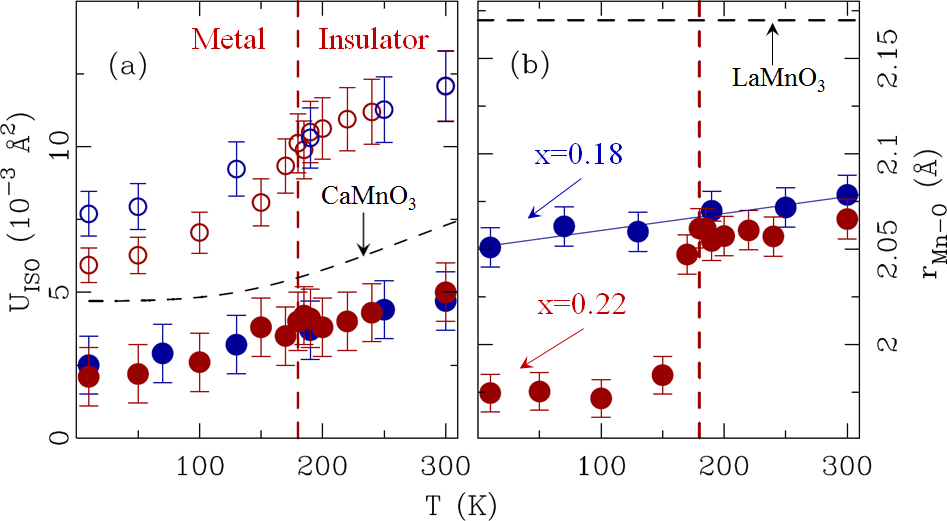}
\caption{(Color online) (a) Comparison of T-evolution of uncorrelated ADPs (Rietveld) for $x=0.18$ (open blue symbols), $x=0.22$ (open red symbols), and $x=1.0$ (dashed black line,
reference) compositions. Correlated ADPs obtained from 2-phase refinements (PDF) are shown by solid symbols for $x=0.18$ (blue) and $x=0.22$ (red). (b) Evolution of long Mn$-$O bond with
temperature from short range PDF modeling. Solid blue symbols represent data for $x=0.18$ (no MIT), solid red symbols represent data for $x=0.22$ (MIT at $\sim$180~K, indicated by vertical dashed red line). Horizontal dashed black line marks long Mn$-$O bond for $x=0$ (reference) composition. Solid blue line through $x=0.18$ points is guide to the eye.
}
  \label{fig;longShort-2phase}
\end{figure}

Given the strong electron-phonon coupling and relatively large size of underlying structural distortions in the PI phase, it is sensible
to utilize structural signatures of the charge localized and delocalized states to explore the existence and to
quantify PS in \lcmo.
In fact, there was one such attempt in the past from earlier \lcmo\ PDF experiments~\cite{proff;apa01i}. In that study, a two-phase model based
on the local structures of the FM and PI phases was used to refine the experimental PDFs quantitatively. The model involved
\cmo$-$like undistorted structural phase and \lmo$-$like distorted structural phase with a distortion size associated with the PI component of
the order of 0.23~\AA\ (defined as difference between the long and short Mn$-$O bond) such as seen in the undoped La-endmember.
Based on the results, a claim was made of the observation of the co-existence
of both phases over a wide temperature range. The fits resulted in approximately 10\% of the localized JT phase (PI) being present even at the lowest
temperature measured (20~K), whereas at room temperature nearly half of the sample remained in the delocalized (FM) phase. However, we note that
in this approach to fitting data the quantification of the phase fraction is highly sensitive to the size of the distortion in the distorted PI phase. At the time of that work, the small polaron model was widely accepted for the insulating phase. In this model, the magnitude of the distortion associated with the charge localized Mn$^{3+}$ sites is expected to be as
large as that in the \lmo\ endmember~\cite{booth;prl98,louca;prb97,billi;prb00}. However a later,  higher resolution, neutron PDF
study~\cite{bozin;prl07} of the insulating regime of \lcmo\ revealed that the {\it local} JT-distortion amplitude in fact decreases dramatically with the
increased Ca-content, contrary to the small polaron model assumption. This observation calls for the early PDF report of temperature evolution of the
phase fraction to be revisited and the phase fraction assignment to be re-examined by a model that uses a more realistic assignment of the underlying
local structure of the metallic and insulating phases.

Here we carried out 2-phase refinements  in a fashion similar to the one described earlier~\cite{proff;apa01i} but with the JT distortions set to those observed in the local structure at 10~K for the $x=0.18$ composition. The structure of the undistorted FM phase was set to that obtained by Rietveld at ~10~K for $x=0.22$ composition. This choice is not optimal as both the average and
local structure change with changing the Ca content. However, the changes occurring between $x=0.18$ and $x=0.22$ are much smaller than those between $x=0$ and the doped
compositions used~\cite{bozin;prl07}. We believe that this assumption is therefore more realistic than the one used in the earlier work~\cite{proff;apa01i}.
In the 2-phase refinements all structural parameters were kept fixed. The only quantities that were allowed to
vary were the phase fraction and isotropic correlated ADPs that were constrained to be the same for the two component-phases.
Typical fits at 10~K are shown in Fig.~\ref{fig;fraction}(b) and (c). Evolution of the observed undistorted fraction with
temperature for the two compositions is presented in Fig.~\ref{fig;fraction}(a). No significant contribution of the undistorted phase could be detected in $x=0.18$ sample, whose
undistorted fraction (and conversely distorted fraction) display almost flat temperature dependence within the accuracy of the assessment. Similarly, the $x=0.22$ sample displays
almost fully undistorted component in the metallic regime, and fully distorted component in the insulating regime, both displaying no temperature dependence, except in the immediate
vicinity of the MI transition. Corresponding correlated ADPs obtained from 2-phase fits are shown in Fig.~\ref{fig;longShort-2phase}(a) (solid symbols) and compared to the
uncorrelated values obtained from single phase fits (open symbols). In addition to results for the undistorted fraction from 2-phase refinements that suggest an abrupt change
with no appreciable doping and temperature dependence across the MI transition, changes in uncorrelated ADPs observed in Rietveld refinement, as well as the observed changes in the local long Mn$-$O bond (Fig.~\ref{fig;longShort-2phase}(b)) are also abrupt and localized to the immediate vicinity of the transition. This strongly suggests that there is no observable phase separation in association with the MI transition, and that doping and temperature dependencies are rather flat, in contrast to what is expected if scenario of percolation of one phase in a phase separated system, where small changes in phase-composition over a wide temperature range would be observed.

%
\subsection{High-T behavior: rhombohedral phase}
\label{rhombohedralPhase}
\begin{figure}[tbp]
  \centering
\includegraphics[angle=0, width=0.475 \textwidth]{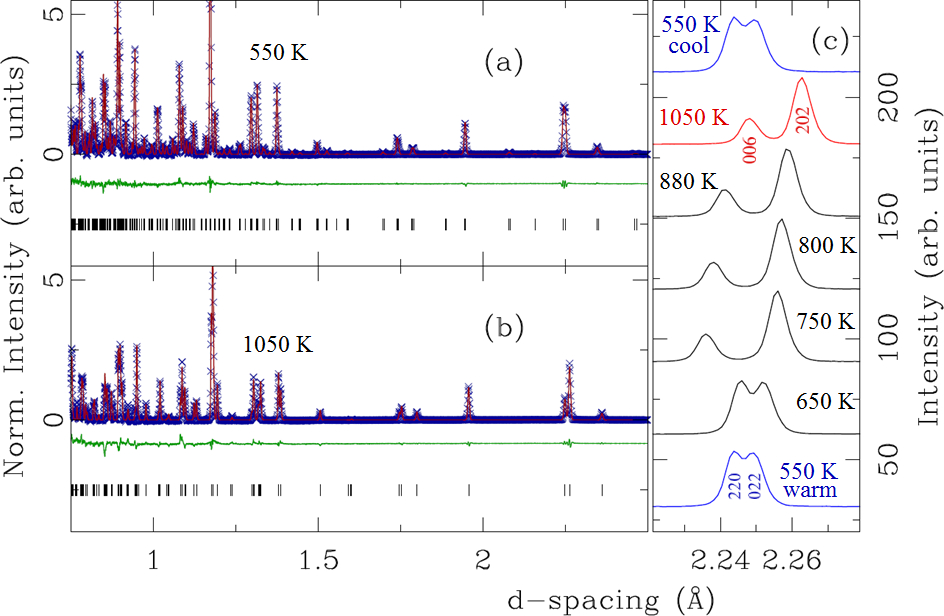}
\caption{(Color online) Rietveld refinements of the average structure models to \lcmo\ $x=0.22$ data at (a) 550~K (\pnma) and (b) 1050~K (\rhm3). Blue symbols represent the data, solid
red lines are the model, and solid green lines are the differences (offset for clarity). Vertical ticks mark reflections. (c) Experimental diffraction patterns over a narrow
range of d-spacing sensitive to O$-$R average structural transition for temperature as indicated. Profiles have been offset for clarity. The bottom and top profiles correspond to the
same 550~K temperature collected on warming and cooling respectively. Oxygen content refined for these two data sets does not change within the experimental uncertainty, verifying that
the sample integrity has been preserved through the heating cycle.
}
  \label{fig;Rietveld-HT}
\end{figure}

%
Finally, we explore the local structure in the high-temperature rhombohedral phase for the $x=0.22$ sample.
Fig.~\ref{fig;Rietveld-HT} shows raw diffraction data and Rietveld refinement fits in this high-temperature region.
Understanding nanosale character of orbital correlations has always been a fundamental challenge in CMR systems in general, and in manganites in particular~\cite{mathu;n97}.
In \lcmo\, a systematic study combining high energy x-ray scattering with elastic and inelastic neutron scattering has been employed recently to explore the character
of the polaron order and dynamics~\cite{lynn;prb07}. It was found that, once established, the nanoscale polaron correlations are only weakly temperature dependent.
These short-range correlations were found to have a static component seen in the elastic scattering channel, indicative of the presence of a glasslike state
 of the polarons, as well as coexisting
dynamic correlations with comparable correlation length, as observed by inelastic neutron scattering. It was further established that the elastic component disappears at
some characteristic higher temperature $T^{*}$, above which the correlations are purely dynamic~\cite{lynn;prb07}.

The character of nanoscale orbital correlations has been studied at temperatures comparable to the JT-transition in \lmo\ endmember, and at temperatures below
the characteristic temperature scale T$^{*}$, within what is known as the pseudocubic phase (still \pnma\ space group)~\cite{lynn;prb07,bozin;prl07}. Explorations
of orbital correlations in the rhombohedral structural phase at temperatures much higher than both T$_{JT}$ in \lmo\ and T$^{*}$ in \lcmo\ have been scarce. An atomic
PDF study explored this temperature regime in the undoped \lmo\ endmember~\cite{qiu;prl05}, and found that the local JT-distortions persist deep into the rhombohedral phase.
Since the PDF method relies indiscriminately on both elastic and inelastic scattering channels, it does not reveal whether the underlying
nanoscale orbital correlations are static or dynamic, although it is reasonable to assume that at such high temperature they are probably purely dynamic.
Studies of the rhombohedral regime for doped \lcmo\ samples have been lacking to date. Recent high-resolution synchrotron x-ray powder diffraction measurement
explored the existence and character of the rhombohedral phase for $x=0.3$ composition~\cite{souza;prb08}. A change of polaronic behavior at the pseudocubic (\pnma) to rhombohedral (\rhm3) structural phase transition (at T$_{S}$) from nonadiabatic in the \pnma\ phase to adiabatic in the \rhm3\ phase was suggested from resistivity measurements. These results in conjunction with neutron and x-ray studies showing polaron correlations in the orthorhombic phase were interpreted as a scenario
in which the structural transition triggers a crossover between distinct electronic states that may be classified as polaron liquid and polaron gas states,
though the density of polarons does not change at this transition, so what this picture really means remains unclear.

\begin{figure}[tbp]
  \centering
\includegraphics[angle=0, width=0.35 \textwidth]{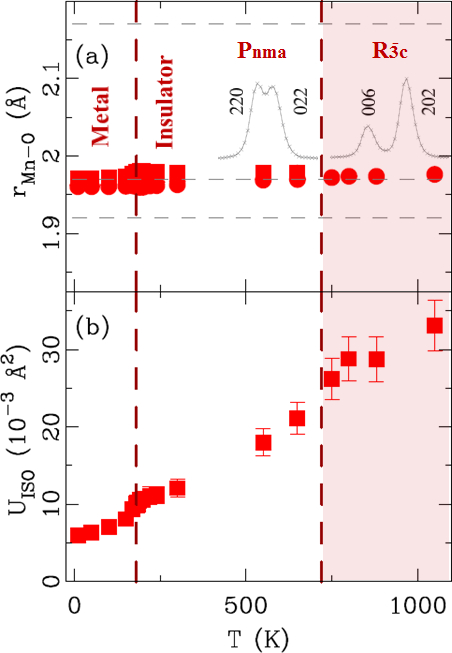}
\caption{(Color online) (a) Temperature evolution of the average Mn$-$O distances of \mo6\ octahedron for $x=0.22$ sample, as obtained from Rietveld refinements across the entire
T-range studied. Vertical dashed red lines at $\sim$~180~K and $\sim$~720~K mark M$-$I and O$-$R transitions, respectively. Vertical dashed lines mark 300~K values for $x=0$ reference.
(b) T-evolution of 8d oxygen (\pnma\ model) ADP as obtained from Rietveld refinements. Inset to (a)
sketches reflections in the diffraction data characteristic for O and R phases, as shown in Fig.~\ref{fig;Rietveld-HT}(c).
}
  \label{fig;averageStructure-HT}
\end{figure}

\begin{figure}[tbp]
  \centering
\includegraphics[angle=0, width=0.35 \textwidth]{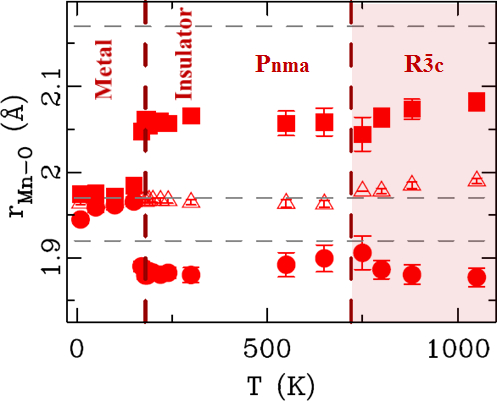}
\caption{(Color online) Temperature evolution of the local Mn$-$O distances of \mo6\ octahedron for $x=0.22$ sample, as obtained from short range PDF refinements. Local distortions
that are onset upon entering the insulating state persist also in the rhombohedral phase, and are seen up to 1050~K - the highest temperature studied. Vertical dashed red lines mark T$_{MI}$ and T$_{S}$.
}
  \label{fig;localStructure-HT}
\end{figure}

In order to shed more light on this part of the phase diagram, we explored the evolution of the average and local atomic structure for $x=0.22$ across the high temperature
transition and deep in the \rhm3\ phase. At this composition T$_{S}$ is $\sim$720~K. We establish the average structural behavior through Rietveld refinements (Fig.~\ref{fig;Rietveld-HT}(a) and (b)), and demonstrate that for the high temperature range studied the sample indeed exhibits the transition where expected.
The
structural phase transition is clear in the raw diffraction data shown in Fig.~\ref{fig;Rietveld-HT}(c), and the respective Rietveld fits do
confirm this (Fig.~\ref{fig;Rietveld-HT}(a), (b)), with data at and above 750~K being in the rhombohedral \rhm3\ spacegroup.
In this spacegroup only a single Mn$-$O bond distance is allowed, and the temperature dependence of the refined value is shown in Fig.~\ref{fig;averageStructure-HT}(a).
It extends continuously from the value in the \pnma\ model at lower temperature.  In that model
three distinct distances are allowed; however they refine to almost the same, single value at higher temperatures indicating that on average
the MnO$_6$ octahedra are already approximately regular even before entering the rhombohedral phase.  However, close examination of the ADPs (Fig.~\ref{fig;averageStructure-HT}(b))
shows that in addition to an anomalous jump in 8d$-$O ADP temperature dependence at T$_{MI}$,
indicative of the onset of local JT-distortions in the insulating phase, there is
also a hint of an additional jump, albeit smaller in size, at around T$_{S}$, similar to what was observed in \lmo~\cite{qiu;prl05}.
As discussed earlier, the system below T$_{S}$ already contains disordered local JT distortions, so this is quite suggestive that the local JT distortions are surviving all the way into the rhombohedral phase, and indeed all the way to our highest temperature data-point at 1050~K.  This is confirmed in the low-$r$ region of the measured PDFs.  Thermal motion of the atoms broaden the PDF peaks making it difficult to resolve clearly the peaks and shoulders directly, but low-$r$ refinements of the PDF clearly prefer distorted MnO$_6$ octahedra extending to the highest temperatures, as shown in Fig.~\ref{fig;localStructure-HT}.

\begin{figure}[tbp]
  \centering
\includegraphics[angle=0, width=0.35 \textwidth]{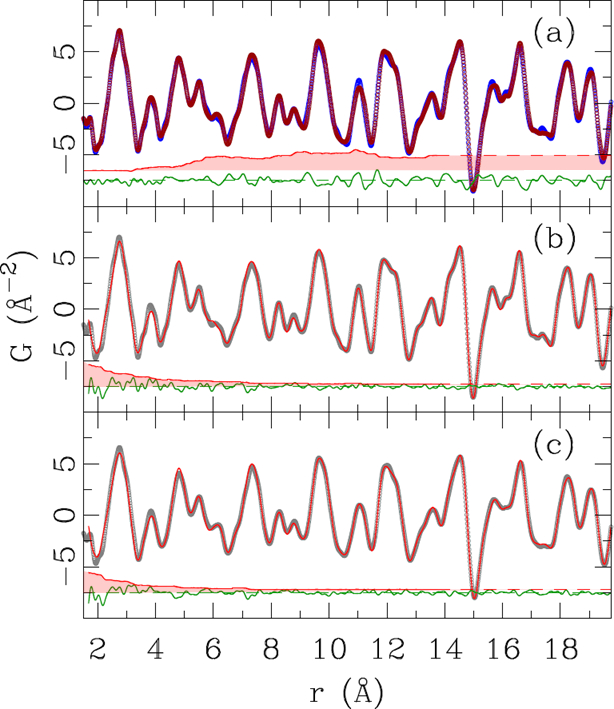}
\caption{(Color online) (a) Comparison of experimental PDF data for $x=0.22$ at 650~K (\pnma\ phase, open blue symbols) and 750~K (\rhm3\ phase, open red symbols). (b)
and (c) show comparison of the data (open gray symbols) and the \pnma\ model based on intermediate range structure parameters (solid red line) for 650~K and 850~K,
respectively. In all panels difference curve (solid green line) is offset for clarity. Light-red solid lines above the differences represent a 4~\AA\ running average of the absolute value of the difference curve, multiplied by 10 and offset for clarity. Shaded areas emphasize the regions of disagreement of the compared PDFs.
}
  \label{fig;localAgreement}
\end{figure}

PDF analysis shows that JT-distortions in $x=0.22$ persist in the \rhm3\ phase
pretty much in the same fashion as they do in \lmo\ endmember~\cite{qiu;prl05}, irrespective
of the fact that the JT-distortion is crystallographically prohibited.
Since the local distortions survive to the highest temperature measured, it is
sensible to ask what does in fact change at T$_{S}$ from the perspective of the
local structural footprint of the electronic/orbital state. As was suggested
earlier~\cite{lynn;prb07}, at a characteristic temperature T$^{*}$ which
extends to T$_{JT}$ for $x=0$ composition (\lmo) there is a change from
presumably polaron glass (static polarons) below T$^{*}$ to a polaron liquid (dynamic polarons)
above T$^{*}$. Since clear JT-distortion
signatures persist still above T$_{S}$, we suggest that at this point there could be a change in the correlations of the dynamic polarons. At such high temperatures, however,
the $r$-dependent PDF fitting protocol is not so stable due to the thermal broadening of the PDF signal and resulting loss in information in the signal.  Instead, we explored an alternative approach (Fig.~\ref{fig;localAgreement}). First, we compare
experimental PDFs for $x=0.22$ at the closest available temperatures bracing the T$_{S}$: 650~K data
correspond to \pnma\ phase, whereas 750~K data correspond to \rhm3\ phase, as can be verified from
inspecting the respective diffraction patterns shown in Fig.~\ref{fig;Rietveld-HT}(c). This
comparison is shown in Fig.~\ref{fig;localAgreement}(a). The difference in the PDFs above and below the transition is captured in the difference curve that is plotted in green below the PDFs. This indicates
that the local structure does not change over the lengthscale of $\sim$5-6~\AA.
Further insights can be gained from assessing how well the model that describes the intermediate range
PDF data explains the data at low-$r$~\cite{bozin;sr14}. To achieve this, a simple two-step procedure was used: in the first step,
the PDF data were fit over $10$~\AA\ $< r < 20$~\AA\ range; in the second step the structure parameters obtained from the first
step were adopted and kept fixed in a refinement of the PDF over $1.7$~\AA\ $< r < 20$~\AA\
range, with model parameters that describe correlated motion of nearest neighbors~\cite{egami;b;utbp12} being the only variables. The procedure was applied to 650~K data (below T$_{S}$)
and 850~K data (above T$_{S}$).
The resulting fits are shown in Fig.~\ref{fig;localAgreement}(b) and (c). The difference curves between the data and the
model display behavior inverse of that seen in Fig.~\ref{fig;localAgreement}(a): agreement is good in the high $r$ region, but poorer
at low $r$, with observable disagreement below $\sim$6~\AA. This not only confirms the existence of local JT-distortions in the
high temperature regime, but also suggests that the range of correlations of the local JT-distortions does not change appreciably on crossing T$_{S}$.
These observations are in line with previous observations of correlated domain size smoothly changing across T$_{S}$ in \lmo\ endmember~\cite{qiu;prl05},
and also consistent with the change of octahedral rotational degrees of freedom.

The results of this study suggest that there does not appear to be a broad range of phase coexistence in this material through the MI transition as has been widely thought previously. In contrast, local JT distortions are rapidly suppressed in the metallic phase ($x=0.22$ at low temperature) on entering both as a function of temperature at
fixed doping, or as a function of doping at fixed temperature.  These results suggest that the percolative picture for the phase transition is unlikely in our samples.  The results are reconciled with earlier PDF studies that suggested otherwise. These studies~\cite{billi;prb00,proff;apa01i} assumed a small-polaron model where the low-$r$ region of the PDF could be rationalized as a mixture of fully JT distorted octahedra and undistorted octahedra.  More recent data show that the amplitude of the local JT distortion actually decreases with doping~\cite{bozin;prl07} and a simple small polaron scenario is not correct.  By taking this into account in the current analysis we obtain a rather different view of the evolution of the metallic and insulating phase fractions with temperature and doping in \lcmo .  The behavior is different from cubic manganite systems with smaller $A$-site ions such as Pr$_{1-x}$Ca$_{x}$MnO$_{3}$\cite{jirak;jmmm85,elova;jpcm14} where phase coexistence has been clearly established on a longer length-scale.  On the other hand, the behavior of \lcmo\ and La$_{1-x}$Sr$_{x}$MnO$_{3}$ systems are likely to be similar to each other, despite the suggestion of a percolative transition in that system from early PDF data~\cite{louca;prb97}.  The persistent feature that was attributed to the Mn$-$O long bond in that study may have been the feature that we attribute to a termination effect in the current work since it is longer in the doped compounds than the  Mn$-$O long-bond in undoped LaMnO$_{3}$ and not shorter as we now know to be the case.  To understand these systems properly, it would be helpful to revisit the La$_{1-x}$Sr$_{x}$MnO$_{3}$ system using modern experimental and analysis protocols.  The behavior of \lcmo\ at higher dopings could also be understood better, where competition between the metallic and ordered-polaronic charge-ordered phases changes the energy balances and therefore phase coexistence behavior~\cite{wu;prl01,loudo;n02,tao;prl05,tao;pnas11}.

\section{Conclusion}
\label{summary}

In summary, we have revisited the metal-insulator transition in doped \lcmo\ ($x=0.18, 0.22$) with a careful neutron PDF study of the local structure.  Distorted MnO$_6$ octahedra are clearly visible in the local structure wherever the samples are in the polaronic insulating phase.  The local JT distortions disappear in the metallic phase. This is in contrast to the observations of the average crystal structure probed either by Rietveld refinement, or refinements of the PDFs carried out over intermediate ranges of $r$ up to 40~\AA.  Both the correlation length of any ordered JT distortion, and also the phase fraction of metallic and insulating material, have been studied as a function of temperature and doping across the MI transition.  We also establish that the local JT distortions persist into the high-temperature rhombohedral phase of the material.
The results of the search for phase coexistence show
that the undistorted fraction changes abruptly across the MI transition, and the phase fraction stays rather uniform at both sides of the phase line. These observations do not support the
percolative scenario for the MI transition in \lcmo\, and bring the ubiquity of phase separation in the broad class of manganites to question. On the other hand, local
distortions of \mo6\ octahedra are found to persist across the pseudocubic to rhombohedral phase transition and deep into the rhombohedral phase where they are
crystallographically prohibited.

%
%
\begin{acknowledgments}
Work at Brookhaven National Laboratory was supported by US DOE, Office of Science, Office of Basic Energy
Sciences (DOE-BES) under contract DE-SC00112704.
Work in the Materials Science Division of Argonne National Laboratory (sample preparation and characterization) was sponsored by the U.S. Department of Energy Office of Science, Basic Energy Sciences, Materials Science and Engineering Division.
Neutron PDF experiments were carried out on NPDF at
LANSCE, funded by DOE BES; Los Alamos National Laboratory is operated by Los Alamos National Security LLC under
contract No. DE-AC52-06NA25396. ESB gratefully acknowledges T. E. Proffen and J. Siewenie for assistance with
the NPDF measurements.
\end{acknowledgments}

%
%


\begin{thebibliography}{58}
\expandafter\ifx\csname natexlab\endcsname\relax\def\natexlab#1{#1}\fi
\expandafter\ifx\csname bibnamefont\endcsname\relax
  \def\bibnamefont#1{#1}\fi
\expandafter\ifx\csname bibfnamefont\endcsname\relax
  \def\bibfnamefont#1{#1}\fi
\expandafter\ifx\csname citenamefont\endcsname\relax
  \def\citenamefont#1{#1}\fi
\expandafter\ifx\csname url\endcsname\relax
  \def\url#1{\texttt{#1}}\fi
\expandafter\ifx\csname urlprefix\endcsname\relax\def\urlprefix{URL }\fi
\providecommand{\bibinfo}[2]{#2}
\providecommand{\eprint}[2][]{\url{#2}}

\bibitem[{\citenamefont{Millis}(1998)}]{milli;n98}
\bibinfo{author}{\bibfnamefont{A.~J.} \bibnamefont{Millis}},
  \bibinfo{journal}{Nature} \textbf{\bibinfo{volume}{392}},
  \bibinfo{pages}{147} (\bibinfo{year}{1998}).

\bibitem[{\citenamefont{Yunoki et~al.}(1998)\citenamefont{Yunoki, Moreo,
  Furukawa, and Dagotto}}]{yuno;prl98}
\bibinfo{author}{\bibfnamefont{S.}~\bibnamefont{Yunoki}},
  \bibinfo{author}{\bibfnamefont{A.}~\bibnamefont{Moreo}},
  \bibinfo{author}{\bibfnamefont{N.}~\bibnamefont{Furukawa}}, \bibnamefont{and}
  \bibinfo{author}{\bibfnamefont{E.}~\bibnamefont{Dagotto}},
  \bibinfo{journal}{Phys. Rev. Lett.} \textbf{\bibinfo{volume}{81}},
  \bibinfo{pages}{5612} (\bibinfo{year}{1998}).

\bibitem[{\citenamefont{Zhou and Goodenough}(2003)}]{zhou;prb03a}
\bibinfo{author}{\bibfnamefont{J.-S.} \bibnamefont{Zhou}} \bibnamefont{and}
  \bibinfo{author}{\bibfnamefont{J.~B.} \bibnamefont{Goodenough}},
  \bibinfo{journal}{Phys. Rev. B} \textbf{\bibinfo{volume}{68}},
  \bibinfo{eid}{144406} (pages~\bibinfo{numpages}{6}) (\bibinfo{year}{2003}),
  \urlprefix\url{http://link.aps.org/abstract/PRB/v68/e144406}.

\bibitem[{\citenamefont{Dagotto}(2005)}]{dagot;s05}
\bibinfo{author}{\bibfnamefont{E.}~\bibnamefont{Dagotto}},
  \bibinfo{journal}{Science} \textbf{\bibinfo{volume}{309}},
  \bibinfo{pages}{257} (\bibinfo{year}{2005}).

\bibitem[{\citenamefont{Bebenin}(2011)}]{beben;pmm11}
\bibinfo{author}{\bibfnamefont{N.}~\bibnamefont{Bebenin}},
  \bibinfo{journal}{Phys. Met. Metallogr.} \textbf{\bibinfo{volume}{111}},
  \bibinfo{pages}{236} (\bibinfo{year}{2011}).

\bibitem[{\citenamefont{Uehara et~al.}(1999)\citenamefont{Uehara, Mori, Chen,
  and {S.-W. Cheong}}}]{uehar;n99}
\bibinfo{author}{\bibfnamefont{M.}~\bibnamefont{Uehara}},
  \bibinfo{author}{\bibfnamefont{S.}~\bibnamefont{Mori}},
  \bibinfo{author}{\bibfnamefont{C.~H.} \bibnamefont{Chen}}, \bibnamefont{and}
  \bibinfo{author}{\bibnamefont{{S.-W. Cheong}}}, \bibinfo{journal}{Nature}
  \textbf{\bibinfo{volume}{399}}, \bibinfo{pages}{560} (\bibinfo{year}{1999}).

\bibitem[{\citenamefont{Dagotto et~al.}(2003)\citenamefont{Dagotto, Burgy, and
  Moreo}}]{dagot;ssc03}
\bibinfo{author}{\bibfnamefont{E.}~\bibnamefont{Dagotto}},
  \bibinfo{author}{\bibfnamefont{J.}~\bibnamefont{Burgy}}, \bibnamefont{and}
  \bibinfo{author}{\bibfnamefont{A.}~\bibnamefont{Moreo}},
  \bibinfo{journal}{Solid State Commun.} \textbf{\bibinfo{volume}{126}},
  \bibinfo{pages}{9} (\bibinfo{year}{2003}).

\bibitem[{\citenamefont{Burgy et~al.}(2004)\citenamefont{Burgy, Moreo, and
  Dagotto}}]{burgy;prl04}
\bibinfo{author}{\bibfnamefont{J.}~\bibnamefont{Burgy}},
  \bibinfo{author}{\bibfnamefont{A.}~\bibnamefont{Moreo}}, \bibnamefont{and}
  \bibinfo{author}{\bibfnamefont{E.}~\bibnamefont{Dagotto}},
  \bibinfo{journal}{Phys. Rev. Lett.} \textbf{\bibinfo{volume}{92}},
  \bibinfo{pages}{097202} (\bibinfo{year}{2004}).

\bibitem[{\citenamefont{Dagotto et~al.}(2008)\citenamefont{Dagotto, Yunoki,
  Sen, Alvarez, and Moreo}}]{dagot;jpcm08}
\bibinfo{author}{\bibfnamefont{E.}~\bibnamefont{Dagotto}},
  \bibinfo{author}{\bibfnamefont{S.}~\bibnamefont{Yunoki}},
  \bibinfo{author}{\bibfnamefont{C.}~\bibnamefont{Sen}},
  \bibinfo{author}{\bibfnamefont{G.}~\bibnamefont{Alvarez}}, \bibnamefont{and}
  \bibinfo{author}{\bibfnamefont{A.}~\bibnamefont{Moreo}}, \bibinfo{journal}{J.
  Phys.: Condens. Mat.} \textbf{\bibinfo{volume}{20}}, \bibinfo{pages}{434224}
  (\bibinfo{year}{2008}).

\bibitem[{\citenamefont{Singh et~al.}(2012)\citenamefont{Singh, Zhang,
  Rajapitamahuni, Devries, and Hong}}]{zhang;physs12}
\bibinfo{author}{\bibfnamefont{V.~R.} \bibnamefont{Singh}},
  \bibinfo{author}{\bibfnamefont{L.}~\bibnamefont{Zhang}},
  \bibinfo{author}{\bibfnamefont{A.~K.} \bibnamefont{Rajapitamahuni}},
  \bibinfo{author}{\bibfnamefont{N.}~\bibnamefont{Devries}}, \bibnamefont{and}
  \bibinfo{author}{\bibfnamefont{X.}~\bibnamefont{Hong}},
  \bibinfo{journal}{Phys. Status Solidi} \textbf{\bibinfo{volume}{9}},
  \bibinfo{pages}{109} (\bibinfo{year}{2012}).

\bibitem[{\citenamefont{Kuberkar et~al.}(2012)\citenamefont{Kuberkar, Doshi,
  Solanki, Khachar, Vagadia, Ravalia, and Ganesan}}]{kuber;ass12}
\bibinfo{author}{\bibfnamefont{D.~G.} \bibnamefont{Kuberkar}},
  \bibinfo{author}{\bibfnamefont{R.~R.} \bibnamefont{Doshi}},
  \bibinfo{author}{\bibfnamefont{P.~S.} \bibnamefont{Solanki}},
  \bibinfo{author}{\bibfnamefont{U.}~\bibnamefont{Khachar}},
  \bibinfo{author}{\bibfnamefont{M.}~\bibnamefont{Vagadia}},
  \bibinfo{author}{\bibfnamefont{A.}~\bibnamefont{Ravalia}}, \bibnamefont{and}
  \bibinfo{author}{\bibfnamefont{V.}~\bibnamefont{Ganesan}},
  \bibinfo{journal}{Appl. Surf. Sci.} \textbf{\bibinfo{volume}{258}},
  \bibinfo{pages}{9041} (\bibinfo{year}{2012}).

\bibitem[{\citenamefont{Wu and Xie}(2012)}]{wu;pb12}
\bibinfo{author}{\bibfnamefont{Z.~H.} \bibnamefont{Wu}} \bibnamefont{and}
  \bibinfo{author}{\bibfnamefont{H.~Q.} \bibnamefont{Xie}},
  \bibinfo{journal}{Physica B} \textbf{\bibinfo{volume}{407}},
  \bibinfo{pages}{2538} (\bibinfo{year}{2012}).

\bibitem[{\citenamefont{Phong et~al.}(2013)\citenamefont{Phong, Manh, Bau, and
  Lee}}]{phong;jec13}
\bibinfo{author}{\bibfnamefont{P.}~\bibnamefont{Phong}},
  \bibinfo{author}{\bibfnamefont{D.}~\bibnamefont{Manh}},
  \bibinfo{author}{\bibfnamefont{L.}~\bibnamefont{Bau}}, \bibnamefont{and}
  \bibinfo{author}{\bibfnamefont{I.-J.} \bibnamefont{Lee}},
  \bibinfo{journal}{J. Electroceram.} \textbf{\bibinfo{volume}{31}},
  \bibinfo{pages}{364} (\bibinfo{year}{2013}).

\bibitem[{\citenamefont{Fath et~al.}(1999)\citenamefont{Fath, Freisem,
  Menovsky, Tomioka, Aarts, and Mydosh}}]{fath;s99}
\bibinfo{author}{\bibfnamefont{M.}~\bibnamefont{Fath}},
  \bibinfo{author}{\bibfnamefont{S.}~\bibnamefont{Freisem}},
  \bibinfo{author}{\bibfnamefont{A.~A.} \bibnamefont{Menovsky}},
  \bibinfo{author}{\bibfnamefont{Y.}~\bibnamefont{Tomioka}},
  \bibinfo{author}{\bibfnamefont{J.}~\bibnamefont{Aarts}}, \bibnamefont{and}
  \bibinfo{author}{\bibfnamefont{J.~A.} \bibnamefont{Mydosh}},
  \bibinfo{journal}{Science} \textbf{\bibinfo{volume}{285}},
  \bibinfo{pages}{1540} (\bibinfo{year}{1999}).

\bibitem[{\citenamefont{de~Teresa et~al.}(1997)\citenamefont{de~Teresa, Ibarra,
  Algarabel, Ritter, Marquina, Blasco, Garcia, del Moral, and
  z.~Arnold}}]{teres;n97}
\bibinfo{author}{\bibfnamefont{J.~M.} \bibnamefont{de~Teresa}},
  \bibinfo{author}{\bibfnamefont{M.~R.} \bibnamefont{Ibarra}},
  \bibinfo{author}{\bibfnamefont{P.~A.} \bibnamefont{Algarabel}},
  \bibinfo{author}{\bibfnamefont{C.}~\bibnamefont{Ritter}},
  \bibinfo{author}{\bibfnamefont{C.}~\bibnamefont{Marquina}},
  \bibinfo{author}{\bibfnamefont{J.}~\bibnamefont{Blasco}},
  \bibinfo{author}{\bibfnamefont{J.}~\bibnamefont{Garcia}},
  \bibinfo{author}{\bibfnamefont{A.}~\bibnamefont{del Moral}},
  \bibnamefont{and} \bibinfo{author}{\bibnamefont{z.~Arnold}},
  \bibinfo{journal}{Nature} \textbf{\bibinfo{volume}{386}},
  \bibinfo{pages}{256} (\bibinfo{year}{1997}).

\bibitem[{\citenamefont{Billinge et~al.}(2000)\citenamefont{Billinge, Proffen,
  Petkov, Sarrao, and Kycia}}]{billi;prb00}
\bibinfo{author}{\bibfnamefont{S.~J.~L.} \bibnamefont{Billinge}},
  \bibinfo{author}{\bibfnamefont{T.}~\bibnamefont{Proffen}},
  \bibinfo{author}{\bibfnamefont{V.}~\bibnamefont{Petkov}},
  \bibinfo{author}{\bibfnamefont{J.~L.} \bibnamefont{Sarrao}},
  \bibnamefont{and} \bibinfo{author}{\bibfnamefont{S.}~\bibnamefont{Kycia}},
  \bibinfo{journal}{Phys. Rev. B} \textbf{\bibinfo{volume}{62}},
  \bibinfo{pages}{1203} (\bibinfo{year}{2000}).

\bibitem[{\citenamefont{Allodi et~al.}(1997)\citenamefont{Allodi, De~Renzi,
  Guidi, Licci, and Pieper}}]{allod;prb97}
\bibinfo{author}{\bibfnamefont{G.}~\bibnamefont{Allodi}},
  \bibinfo{author}{\bibfnamefont{R.}~\bibnamefont{De~Renzi}},
  \bibinfo{author}{\bibfnamefont{G.}~\bibnamefont{Guidi}},
  \bibinfo{author}{\bibfnamefont{F.}~\bibnamefont{Licci}}, \bibnamefont{and}
  \bibinfo{author}{\bibfnamefont{M.~W.} \bibnamefont{Pieper}},
  \bibinfo{journal}{Phys. Rev. B} \textbf{\bibinfo{volume}{56}},
  \bibinfo{pages}{6036} (\bibinfo{year}{1997}).

\bibitem[{\citenamefont{Kumar et~al.}(2014)\citenamefont{Kumar, Mathieu,
  Nordblad, Ray, Karis, Andersson, and Sarma}}]{kumar;prx14}
\bibinfo{author}{\bibfnamefont{P.~A.} \bibnamefont{Kumar}},
  \bibinfo{author}{\bibfnamefont{R.}~\bibnamefont{Mathieu}},
  \bibinfo{author}{\bibfnamefont{P.}~\bibnamefont{Nordblad}},
  \bibinfo{author}{\bibfnamefont{S.}~\bibnamefont{Ray}},
  \bibinfo{author}{\bibfnamefont{O.}~\bibnamefont{Karis}},
  \bibinfo{author}{\bibfnamefont{G.}~\bibnamefont{Andersson}},
  \bibnamefont{and} \bibinfo{author}{\bibfnamefont{D.~D.} \bibnamefont{Sarma}},
  \bibinfo{journal}{Phys. Rev. X} \textbf{\bibinfo{volume}{4}},
  \bibinfo{pages}{011037} (\bibinfo{year}{2014}).

\bibitem[{\citenamefont{Mitra et~al.}(2005)\citenamefont{Mitra, Paranjape,
  Raychaudhuri, Mathur, and Blamire}}]{mitra;prb05}
\bibinfo{author}{\bibfnamefont{J.}~\bibnamefont{Mitra}},
  \bibinfo{author}{\bibfnamefont{M.}~\bibnamefont{Paranjape}},
  \bibinfo{author}{\bibfnamefont{A.~K.} \bibnamefont{Raychaudhuri}},
  \bibinfo{author}{\bibfnamefont{N.~D.} \bibnamefont{Mathur}},
  \bibnamefont{and} \bibinfo{author}{\bibfnamefont{M.~G.}
  \bibnamefont{Blamire}}, \bibinfo{journal}{Phys. Rev. B}
  \textbf{\bibinfo{volume}{71}}, \bibinfo{pages}{094426}
  (\bibinfo{year}{2005}).

\bibitem[{\citenamefont{Tyson et~al.}(1996)\citenamefont{Tyson, {J. Mustre de
  Leon}, Conradson, Bishop, Neumeier, R\"oder, and Zang}}]{tyson;prb96}
\bibinfo{author}{\bibfnamefont{T.~A.} \bibnamefont{Tyson}},
  \bibinfo{author}{\bibnamefont{{J. Mustre de Leon}}},
  \bibinfo{author}{\bibfnamefont{S.~D.} \bibnamefont{Conradson}},
  \bibinfo{author}{\bibfnamefont{A.~R.} \bibnamefont{Bishop}},
  \bibinfo{author}{\bibfnamefont{J.~J.} \bibnamefont{Neumeier}},
  \bibinfo{author}{\bibfnamefont{H.}~\bibnamefont{R\"oder}}, \bibnamefont{and}
  \bibinfo{author}{\bibfnamefont{J.}~\bibnamefont{Zang}},
  \bibinfo{journal}{Phys. Rev. B} \textbf{\bibinfo{volume}{53}},
  \bibinfo{pages}{13985} (\bibinfo{year}{1996}).

\bibitem[{\citenamefont{Hundley et~al.}(1995)\citenamefont{Hundley, Hawley,
  Heffner, Jia, Neumeier, Tesmer, Thompson, and Wu}}]{hundl;apl95}
\bibinfo{author}{\bibfnamefont{M.~F.} \bibnamefont{Hundley}},
  \bibinfo{author}{\bibfnamefont{M.}~\bibnamefont{Hawley}},
  \bibinfo{author}{\bibfnamefont{R.~H.} \bibnamefont{Heffner}},
  \bibinfo{author}{\bibfnamefont{Q.~X.} \bibnamefont{Jia}},
  \bibinfo{author}{\bibfnamefont{J.~J.} \bibnamefont{Neumeier}},
  \bibinfo{author}{\bibfnamefont{J.}~\bibnamefont{Tesmer}},
  \bibinfo{author}{\bibfnamefont{J.~D.} \bibnamefont{Thompson}},
  \bibnamefont{and} \bibinfo{author}{\bibfnamefont{X.~D.} \bibnamefont{Wu}},
  \bibinfo{journal}{Appl. Phys. Lett.} \textbf{\bibinfo{volume}{67}},
  \bibinfo{pages}{860} (\bibinfo{year}{1995}).

\bibitem[{\citenamefont{Billinge et~al.}(1996)\citenamefont{Billinge,
  DiFrancesco, Kwei, Neumeier, and Thompson}}]{billi;prl96}
\bibinfo{author}{\bibfnamefont{S.~J.~L.} \bibnamefont{Billinge}},
  \bibinfo{author}{\bibfnamefont{R.~G.} \bibnamefont{DiFrancesco}},
  \bibinfo{author}{\bibfnamefont{G.~H.} \bibnamefont{Kwei}},
  \bibinfo{author}{\bibfnamefont{J.~J.} \bibnamefont{Neumeier}},
  \bibnamefont{and} \bibinfo{author}{\bibfnamefont{J.~D.}
  \bibnamefont{Thompson}}, \bibinfo{journal}{Phys. Rev. Lett.}
  \textbf{\bibinfo{volume}{77}}, \bibinfo{pages}{715} (\bibinfo{year}{1996}).

\bibitem[{\citenamefont{Qiu et~al.}(2005)\citenamefont{Qiu, Proffen, Mitchell,
  and Billinge}}]{qiu;prl05}
\bibinfo{author}{\bibfnamefont{X.}~\bibnamefont{Qiu}},
  \bibinfo{author}{\bibfnamefont{T.}~\bibnamefont{Proffen}},
  \bibinfo{author}{\bibfnamefont{J.~F.} \bibnamefont{Mitchell}},
  \bibnamefont{and} \bibinfo{author}{\bibfnamefont{S.~J.~L.}
  \bibnamefont{Billinge}}, \bibinfo{journal}{Phys. Rev. Lett.}
  \textbf{\bibinfo{volume}{94}}, \bibinfo{pages}{177203}
  (\bibinfo{year}{2005}).

\bibitem[{\citenamefont{Sartbaeva et~al.}(2007)\citenamefont{Sartbaeva, Wells,
  Thorpe, {Bo\v zin}, and Billinge}}]{sartb;prl07}
\bibinfo{author}{\bibfnamefont{A.}~\bibnamefont{Sartbaeva}},
  \bibinfo{author}{\bibfnamefont{S.~A.} \bibnamefont{Wells}},
  \bibinfo{author}{\bibfnamefont{M.~F.} \bibnamefont{Thorpe}},
  \bibinfo{author}{\bibfnamefont{E.~S.} \bibnamefont{{Bo\v zin}}},
  \bibnamefont{and} \bibinfo{author}{\bibfnamefont{S.~J.~L.}
  \bibnamefont{Billinge}}, \bibinfo{journal}{Phys. Rev. Lett.}
  \textbf{\bibinfo{volume}{99}}, \bibinfo{pages}{155503}
  (\bibinfo{year}{2007}).

\bibitem[{\citenamefont{Kiryukhin}(2004)}]{kiryu;njp04}
\bibinfo{author}{\bibfnamefont{V.}~\bibnamefont{Kiryukhin}},
  \bibinfo{journal}{New J. Phys.} \textbf{\bibinfo{volume}{6}},
  \bibinfo{pages}{155} (\bibinfo{year}{2004}).

\bibitem[{\citenamefont{Bo{\v z}in et~al.}(2007)\citenamefont{Bo{\v z}in,
  Schmidt, {DeConinck}, Paglia, Mitchell, Chatterji, Radaelli, Proffen, and
  Billinge}}]{bozin;prl07}
\bibinfo{author}{\bibfnamefont{E.~S.} \bibnamefont{Bo{\v z}in}},
  \bibinfo{author}{\bibfnamefont{M.}~\bibnamefont{Schmidt}},
  \bibinfo{author}{\bibfnamefont{A.~J.} \bibnamefont{{DeConinck}}},
  \bibinfo{author}{\bibfnamefont{G.}~\bibnamefont{Paglia}},
  \bibinfo{author}{\bibfnamefont{J.~F.} \bibnamefont{Mitchell}},
  \bibinfo{author}{\bibfnamefont{T.}~\bibnamefont{Chatterji}},
  \bibinfo{author}{\bibfnamefont{P.~G.} \bibnamefont{Radaelli}},
  \bibinfo{author}{\bibfnamefont{T.}~\bibnamefont{Proffen}}, \bibnamefont{and}
  \bibinfo{author}{\bibfnamefont{S.~J.~L.} \bibnamefont{Billinge}},
  \bibinfo{journal}{Phys. Rev. Lett.} \textbf{\bibinfo{volume}{98}},
  \bibinfo{pages}{137203} (\bibinfo{year}{2007}),
  \urlprefix\url{http://journals.aps.org/prl/abstract/10.1103/PhysRevLett.98.137203}.

\bibitem[{\citenamefont{Chatterji et~al.}(2003)\citenamefont{Chatterji, Fauth,
  Ouladdiaf, Mandal, and Ghosh}}]{chatt;prb03}
\bibinfo{author}{\bibfnamefont{T.}~\bibnamefont{Chatterji}},
  \bibinfo{author}{\bibfnamefont{F.}~\bibnamefont{Fauth}},
  \bibinfo{author}{\bibfnamefont{B.}~\bibnamefont{Ouladdiaf}},
  \bibinfo{author}{\bibfnamefont{P.}~\bibnamefont{Mandal}}, \bibnamefont{and}
  \bibinfo{author}{\bibfnamefont{B.}~\bibnamefont{Ghosh}},
  \bibinfo{journal}{Phys. Rev. B} \textbf{\bibinfo{volume}{68}},
  \bibinfo{eid}{052406} (pages~\bibinfo{numpages}{4}) (\bibinfo{year}{2003}).

\bibitem[{\citenamefont{Dabrowski et~al.}(1989)\citenamefont{Dabrowski,
  Dybzinski, Bukowski, Chmaissem, and Jorgensen}}]{dabro;jssc89}
\bibinfo{author}{\bibfnamefont{B.}~\bibnamefont{Dabrowski}},
  \bibinfo{author}{\bibfnamefont{R.}~\bibnamefont{Dybzinski}},
  \bibinfo{author}{\bibfnamefont{Z.}~\bibnamefont{Bukowski}},
  \bibinfo{author}{\bibfnamefont{O.}~\bibnamefont{Chmaissem}},
  \bibnamefont{and} \bibinfo{author}{\bibfnamefont{J.~D.}
  \bibnamefont{Jorgensen}}, \bibinfo{journal}{J. Solid State Chem.}
  \textbf{\bibinfo{volume}{146}}, \bibinfo{pages}{448} (\bibinfo{year}{1989}).

\bibitem[{\citenamefont{Salamon and Jaime}(2001)}]{salam;rmp01}
\bibinfo{author}{\bibfnamefont{M.~B.} \bibnamefont{Salamon}} \bibnamefont{and}
  \bibinfo{author}{\bibfnamefont{M.}~\bibnamefont{Jaime}},
  \bibinfo{journal}{Rev. Mod. Phys.} \textbf{\bibinfo{volume}{73}},
  \bibinfo{pages}{583} (\bibinfo{year}{2001}).

\bibitem[{\citenamefont{Egami and Billinge}(2012)}]{egami;b;utbp12}
\bibinfo{author}{\bibfnamefont{T.}~\bibnamefont{Egami}} \bibnamefont{and}
  \bibinfo{author}{\bibfnamefont{S.~J.~L.} \bibnamefont{Billinge}},
  \emph{\bibinfo{title}{Underneath the Bragg peaks: structural analysis of
  complex materials}} (\bibinfo{publisher}{Elsevier},
  \bibinfo{address}{Amsterdam}, \bibinfo{year}{2012}), \bibinfo{edition}{2nd}
  ed.,
  \urlprefix\url{http://store.elsevier.com/product.jsp?lid=0\&iid=73\&sid=0\&isbn=9780080971414}.

\bibitem[{\citenamefont{Peterson et~al.}(2000)\citenamefont{Peterson, Gutmann,
  Proffen, and Billinge}}]{peter;jac00}
\bibinfo{author}{\bibfnamefont{P.~F.} \bibnamefont{Peterson}},
  \bibinfo{author}{\bibfnamefont{M.}~\bibnamefont{Gutmann}},
  \bibinfo{author}{\bibfnamefont{T.}~\bibnamefont{Proffen}}, \bibnamefont{and}
  \bibinfo{author}{\bibfnamefont{S.~J.~L.} \bibnamefont{Billinge}},
  \bibinfo{journal}{J. Appl. Crystallogr.} \textbf{\bibinfo{volume}{33}},
  \bibinfo{pages}{1192} (\bibinfo{year}{2000}),
  \urlprefix\url{http://dx.doi.org/10.1107/S0021889800007123}.

\bibitem[{\citenamefont{Rietveld}(1967)}]{rietv;ac67}
\bibinfo{author}{\bibfnamefont{H.~M.} \bibnamefont{Rietveld}},
  \bibinfo{journal}{Acta Crystallogr.} \textbf{\bibinfo{volume}{22}},
  \bibinfo{pages}{151 } (\bibinfo{year}{1967}).

\bibitem[{\citenamefont{Larson and {Von Dreele}}(1987)}]{larso;unpub87}
\bibinfo{author}{\bibfnamefont{A.~C.} \bibnamefont{Larson}} \bibnamefont{and}
  \bibinfo{author}{\bibfnamefont{R.~B.} \bibnamefont{{Von Dreele}}}
  (\bibinfo{year}{1987}), \bibinfo{note}{report No. LAUR-86-748, Los Alamos
  National Laboratory, Los Alamos, NM 87545}.

\bibitem[{\citenamefont{Toby}(2001)}]{toby;jac01}
\bibinfo{author}{\bibfnamefont{B.~H.} \bibnamefont{Toby}}, \bibinfo{journal}{J.
  Appl. Crystallogr.} \textbf{\bibinfo{volume}{34}}, \bibinfo{pages}{201}
  (\bibinfo{year}{2001}).

\bibitem[{\citenamefont{Farrow et~al.}(2007)\citenamefont{Farrow, Juh\'as, Liu,
  Bryndin, {Bo\v zin}, Bloch, Proffen, and Billinge}}]{farro;jpcm07}
\bibinfo{author}{\bibfnamefont{C.~L.} \bibnamefont{Farrow}},
  \bibinfo{author}{\bibfnamefont{P.}~\bibnamefont{Juh\'as}},
  \bibinfo{author}{\bibfnamefont{J.}~\bibnamefont{Liu}},
  \bibinfo{author}{\bibfnamefont{D.}~\bibnamefont{Bryndin}},
  \bibinfo{author}{\bibfnamefont{E.~S.} \bibnamefont{{Bo\v zin}}},
  \bibinfo{author}{\bibfnamefont{J.}~\bibnamefont{Bloch}},
  \bibinfo{author}{\bibfnamefont{T.}~\bibnamefont{Proffen}}, \bibnamefont{and}
  \bibinfo{author}{\bibfnamefont{S.~J.~L.} \bibnamefont{Billinge}},
  \bibinfo{journal}{J. Phys: Condens. Mat.} \textbf{\bibinfo{volume}{19}},
  \bibinfo{pages}{335219} (\bibinfo{year}{2007}),
  \urlprefix\url{http://iopscience.iop.org/0953-8984/19/33/335219/}.

\bibitem[{\citenamefont{Bo\v{z}in et~al.}(2008)\citenamefont{Bo\v{z}in,
  Sartbaeva, Zheng, Wells, Mitchell, Proffen, Thorpe, and
  Billinge}}]{bozin;jpcs08}
\bibinfo{author}{\bibfnamefont{E.~S.} \bibnamefont{Bo\v{z}in}},
  \bibinfo{author}{\bibfnamefont{A.}~\bibnamefont{Sartbaeva}},
  \bibinfo{author}{\bibfnamefont{H.}~\bibnamefont{Zheng}},
  \bibinfo{author}{\bibfnamefont{S.~A.} \bibnamefont{Wells}},
  \bibinfo{author}{\bibfnamefont{J.~F.} \bibnamefont{Mitchell}},
  \bibinfo{author}{\bibfnamefont{T.}~\bibnamefont{Proffen}},
  \bibinfo{author}{\bibfnamefont{M.~F.} \bibnamefont{Thorpe}},
  \bibnamefont{and} \bibinfo{author}{\bibfnamefont{S.~J.~L.}
  \bibnamefont{Billinge}}, \bibinfo{journal}{J. Phys. Chem. Solids}
  \textbf{\bibinfo{volume}{69}}, \bibinfo{pages}{2146 } (\bibinfo{year}{2008}),
  \urlprefix\url{http://www.sciencedirect.com/science/article/pii/S0022369708000917}.

\bibitem[{\citenamefont{Huang et~al.}(1997)\citenamefont{Huang, Santoro, Lynn,
  Erwin, Borchers, Peng, and Greene}}]{huang;prb97}
\bibinfo{author}{\bibfnamefont{Q.}~\bibnamefont{Huang}},
  \bibinfo{author}{\bibfnamefont{A.}~\bibnamefont{Santoro}},
  \bibinfo{author}{\bibfnamefont{J.~W.} \bibnamefont{Lynn}},
  \bibinfo{author}{\bibfnamefont{R.~W.} \bibnamefont{Erwin}},
  \bibinfo{author}{\bibfnamefont{J.~A.} \bibnamefont{Borchers}},
  \bibinfo{author}{\bibfnamefont{J.~L.} \bibnamefont{Peng}}, \bibnamefont{and}
  \bibinfo{author}{\bibfnamefont{R.~L.} \bibnamefont{Greene}},
  \bibinfo{journal}{Phys. Rev. B} \textbf{\bibinfo{volume}{55}},
  \bibinfo{pages}{14987} (\bibinfo{year}{1997}).

\bibitem[{\citenamefont{Souza et~al.}(2008)\citenamefont{Souza, Terashita,
  Granado, Jardim, {N. F. Oliveira Jr.}, and Muccillo}}]{souza;prb08}
\bibinfo{author}{\bibfnamefont{J.~A.} \bibnamefont{Souza}},
  \bibinfo{author}{\bibfnamefont{H.}~\bibnamefont{Terashita}},
  \bibinfo{author}{\bibfnamefont{E.}~\bibnamefont{Granado}},
  \bibinfo{author}{\bibfnamefont{R.~F.} \bibnamefont{Jardim}},
  \bibinfo{author}{\bibnamefont{{N. F. Oliveira Jr.}}}, \bibnamefont{and}
  \bibinfo{author}{\bibfnamefont{R.}~\bibnamefont{Muccillo}},
  \bibinfo{journal}{Phys. Rev. B} \textbf{\bibinfo{volume}{78}},
  \bibinfo{pages}{054411} (\bibinfo{year}{2008}).

\bibitem[{\citenamefont{Chatterji et~al.}(2002)\citenamefont{Chatterji,
  Ouladdiaf, Mandal, Bandyopadhyay, and Ghosh}}]{chatt;prb02}
\bibinfo{author}{\bibfnamefont{T.}~\bibnamefont{Chatterji}},
  \bibinfo{author}{\bibfnamefont{B.}~\bibnamefont{Ouladdiaf}},
  \bibinfo{author}{\bibfnamefont{P.}~\bibnamefont{Mandal}},
  \bibinfo{author}{\bibfnamefont{B.}~\bibnamefont{Bandyopadhyay}},
  \bibnamefont{and} \bibinfo{author}{\bibfnamefont{B.}~\bibnamefont{Ghosh}},
  \bibinfo{journal}{Phys. Rev. B} \textbf{\bibinfo{volume}{66}},
  \bibinfo{pages}{054403/1} (\bibinfo{year}{2002}).

\bibitem[{\citenamefont{Kiryukhin et~al.}(2003)\citenamefont{Kiryukhin, Koo,
  Ishibashi, Hill, and Cheong}}]{kiryu;prb03}
\bibinfo{author}{\bibfnamefont{V.}~\bibnamefont{Kiryukhin}},
  \bibinfo{author}{\bibfnamefont{T.~Y.} \bibnamefont{Koo}},
  \bibinfo{author}{\bibfnamefont{H.}~\bibnamefont{Ishibashi}},
  \bibinfo{author}{\bibfnamefont{J.~P.} \bibnamefont{Hill}}, \bibnamefont{and}
  \bibinfo{author}{\bibfnamefont{S.}~\bibnamefont{Cheong}},
  \bibinfo{journal}{Phys. Rev. B} \textbf{\bibinfo{volume}{67}},
  \bibinfo{pages}{064421} (\bibinfo{year}{2003}).

\bibitem[{\citenamefont{Kiryukhin et~al.}(2004)\citenamefont{Kiryukhin,
  Borissov, Ahn, Huang, Lynn, and Cheong}}]{kiryu;prb04}
\bibinfo{author}{\bibfnamefont{V.}~\bibnamefont{Kiryukhin}},
  \bibinfo{author}{\bibfnamefont{A.}~\bibnamefont{Borissov}},
  \bibinfo{author}{\bibfnamefont{J.~S.} \bibnamefont{Ahn}},
  \bibinfo{author}{\bibfnamefont{Q.}~\bibnamefont{Huang}},
  \bibinfo{author}{\bibfnamefont{J.~W.} \bibnamefont{Lynn}}, \bibnamefont{and}
  \bibinfo{author}{\bibfnamefont{S.}~\bibnamefont{Cheong}},
  \bibinfo{journal}{Phys. Rev. B} \textbf{\bibinfo{volume}{70}},
  \bibinfo{pages}{214424} (\bibinfo{year}{2004}).

\bibitem[{\citenamefont{Radaelli et~al.}(1996)\citenamefont{Radaelli, Marezio,
  Hwang, Cheong, and Batlogg}}]{radae;prb96i}
\bibinfo{author}{\bibfnamefont{P.~G.} \bibnamefont{Radaelli}},
  \bibinfo{author}{\bibfnamefont{M.}~\bibnamefont{Marezio}},
  \bibinfo{author}{\bibfnamefont{H.~Y.} \bibnamefont{Hwang}},
  \bibinfo{author}{\bibfnamefont{S.}~\bibnamefont{Cheong}}, \bibnamefont{and}
  \bibinfo{author}{\bibfnamefont{B.}~\bibnamefont{Batlogg}},
  \bibinfo{journal}{Phys. Rev. B} \textbf{\bibinfo{volume}{54}},
  \bibinfo{pages}{8992} (\bibinfo{year}{1996}).

\bibitem[{\citenamefont{Booth et~al.}(1998)\citenamefont{Booth, Bridges, Kwei,
  Lawrence, Cornelius, and Neumeier}}]{booth;prl98}
\bibinfo{author}{\bibfnamefont{C.~H.} \bibnamefont{Booth}},
  \bibinfo{author}{\bibfnamefont{F.}~\bibnamefont{Bridges}},
  \bibinfo{author}{\bibfnamefont{G.~H.} \bibnamefont{Kwei}},
  \bibinfo{author}{\bibfnamefont{J.~M.} \bibnamefont{Lawrence}},
  \bibinfo{author}{\bibfnamefont{A.~L.} \bibnamefont{Cornelius}},
  \bibnamefont{and} \bibinfo{author}{\bibfnamefont{J.~J.}
  \bibnamefont{Neumeier}}, \bibinfo{journal}{Phys. Rev. Lett.}
  \textbf{\bibinfo{volume}{80}}, \bibinfo{pages}{853} (\bibinfo{year}{1998}).

\bibitem[{\citenamefont{Louca et~al.}(1997)\citenamefont{Louca, Egami, Brosha,
  {R\"{o}der}, and Bishop}}]{louca;prb97}
\bibinfo{author}{\bibfnamefont{D.}~\bibnamefont{Louca}},
  \bibinfo{author}{\bibfnamefont{T.}~\bibnamefont{Egami}},
  \bibinfo{author}{\bibfnamefont{E.~L.} \bibnamefont{Brosha}},
  \bibinfo{author}{\bibfnamefont{H.}~\bibnamefont{{R\"{o}der}}},
  \bibnamefont{and} \bibinfo{author}{\bibfnamefont{A.~R.}
  \bibnamefont{Bishop}}, \bibinfo{journal}{Phys. Rev. B}
  \textbf{\bibinfo{volume}{56}}, \bibinfo{pages}{R8475} (\bibinfo{year}{1997}).

\bibitem[{\citenamefont{Bo\v{z}in et~al.}(2014)\citenamefont{Bo\v{z}in, Knox,
  Juh\'{a}s, Hor, Mitchell, and Billinge}}]{bozin;sr14}
\bibinfo{author}{\bibfnamefont{E.~S.} \bibnamefont{Bo\v{z}in}},
  \bibinfo{author}{\bibfnamefont{K.~R.} \bibnamefont{Knox}},
  \bibinfo{author}{\bibfnamefont{P.}~\bibnamefont{Juh\'{a}s}},
  \bibinfo{author}{\bibfnamefont{Y.~S.} \bibnamefont{Hor}},
  \bibinfo{author}{\bibfnamefont{J.~F.} \bibnamefont{Mitchell}},
  \bibnamefont{and} \bibinfo{author}{\bibfnamefont{S.~J.~L.}
  \bibnamefont{Billinge}}, \bibinfo{journal}{Sci. Rep.}
  \textbf{\bibinfo{volume}{4}}, \bibinfo{pages}{4081} (\bibinfo{year}{2014}),
  \urlprefix\url{http://www.nature.com/srep/2014/140212/srep04081/full/srep04081.html}.

\bibitem[{\citenamefont{Martin et~al.}(1999)\citenamefont{Martin, Maignan,
  Hervieu, and Raveau}}]{marti;prb99}
\bibinfo{author}{\bibfnamefont{C.}~\bibnamefont{Martin}},
  \bibinfo{author}{\bibfnamefont{A.}~\bibnamefont{Maignan}},
  \bibinfo{author}{\bibfnamefont{M.}~\bibnamefont{Hervieu}}, \bibnamefont{and}
  \bibinfo{author}{\bibfnamefont{B.}~\bibnamefont{Raveau}},
  \bibinfo{journal}{Phys. Rev. B} \textbf{\bibinfo{volume}{60}},
  \bibinfo{pages}{12191} (\bibinfo{year}{1999}).

\bibitem[{\citenamefont{Becker et~al.}(2002)\citenamefont{Becker, Streng, Luo,
  Moshnyaga, Damaschke, Shannon, and Samwer}}]{becke;prl02}
\bibinfo{author}{\bibfnamefont{T.}~\bibnamefont{Becker}},
  \bibinfo{author}{\bibfnamefont{C.}~\bibnamefont{Streng}},
  \bibinfo{author}{\bibfnamefont{Y.}~\bibnamefont{Luo}},
  \bibinfo{author}{\bibfnamefont{V.}~\bibnamefont{Moshnyaga}},
  \bibinfo{author}{\bibfnamefont{B.}~\bibnamefont{Damaschke}},
  \bibinfo{author}{\bibfnamefont{N.}~\bibnamefont{Shannon}}, \bibnamefont{and}
  \bibinfo{author}{\bibfnamefont{K.}~\bibnamefont{Samwer}},
  \bibinfo{journal}{Phys. Rev. Lett.} \textbf{\bibinfo{volume}{89}},
  \bibinfo{pages}{237203} (\bibinfo{year}{2002}).

\bibitem[{\citenamefont{Ramesh et~al.}(2009)\citenamefont{Ramesh, Han, Ning,
  Cheng, Sun, and Jayavel}}]{rames;ml09}
\bibinfo{author}{\bibfnamefont{B.~M.} \bibnamefont{Ramesh}},
  \bibinfo{author}{\bibfnamefont{X.}~\bibnamefont{Han}},
  \bibinfo{author}{\bibfnamefont{W.}~\bibnamefont{Ning}},
  \bibinfo{author}{\bibfnamefont{Z.-h.} \bibnamefont{Cheng}},
  \bibinfo{author}{\bibfnamefont{Y.}~\bibnamefont{Sun}}, \bibnamefont{and}
  \bibinfo{author}{\bibfnamefont{R.}~\bibnamefont{Jayavel}},
  \bibinfo{journal}{Mater. Lett.} \textbf{\bibinfo{volume}{63}},
  \bibinfo{pages}{1528} (\bibinfo{year}{2009}).

\bibitem[{\citenamefont{Ward et~al.}(2011)\citenamefont{Ward, Gai, Guo, Yin,
  and Shen}}]{ward;prb11}
\bibinfo{author}{\bibfnamefont{T.~Z.} \bibnamefont{Ward}},
  \bibinfo{author}{\bibfnamefont{Z.}~\bibnamefont{Gai}},
  \bibinfo{author}{\bibfnamefont{H.~W.} \bibnamefont{Guo}},
  \bibinfo{author}{\bibfnamefont{L.~F.} \bibnamefont{Yin}}, \bibnamefont{and}
  \bibinfo{author}{\bibfnamefont{J.}~\bibnamefont{Shen}},
  \bibinfo{journal}{Phys. Rev. B} \textbf{\bibinfo{volume}{83}},
  \bibinfo{pages}{125125} (\bibinfo{year}{2011}).

\bibitem[{\citenamefont{Tao et~al.}(2011)\citenamefont{Tao, Niebieskikwiat,
  Jie, Schofield, Wu, Li, and Zhu}}]{tao;pnas11}
\bibinfo{author}{\bibfnamefont{J.}~\bibnamefont{Tao}},
  \bibinfo{author}{\bibfnamefont{D.}~\bibnamefont{Niebieskikwiat}},
  \bibinfo{author}{\bibfnamefont{Q.}~\bibnamefont{Jie}},
  \bibinfo{author}{\bibfnamefont{M.~A.} \bibnamefont{Schofield}},
  \bibinfo{author}{\bibfnamefont{L.~J.} \bibnamefont{Wu}},
  \bibinfo{author}{\bibfnamefont{Q.}~\bibnamefont{Li}}, \bibnamefont{and}
  \bibinfo{author}{\bibfnamefont{Y.~M.} \bibnamefont{Zhu}},
  \bibinfo{journal}{Proc. Natl. Acad. Sci. USA} \textbf{\bibinfo{volume}{108}},
  \bibinfo{pages}{20941} (\bibinfo{year}{2011}).

\bibitem[{\citenamefont{Loudon et~al.}(2002)\citenamefont{Loudon, Mathur, and
  Midgley}}]{loudo;n02}
\bibinfo{author}{\bibfnamefont{J.~C.} \bibnamefont{Loudon}},
  \bibinfo{author}{\bibfnamefont{N.~D.} \bibnamefont{Mathur}},
  \bibnamefont{and} \bibinfo{author}{\bibfnamefont{P.~A.}
  \bibnamefont{Midgley}}, \bibinfo{journal}{Nature}
  \textbf{\bibinfo{volume}{420}}, \bibinfo{pages}{797} (\bibinfo{year}{2002}).

\bibitem[{\citenamefont{Proffen et~al.}(2002)\citenamefont{Proffen, Egami,
  Billinge, Cheetham, Louca, and Parise}}]{proff;apa01i}
\bibinfo{author}{\bibfnamefont{T.}~\bibnamefont{Proffen}},
  \bibinfo{author}{\bibfnamefont{T.}~\bibnamefont{Egami}},
  \bibinfo{author}{\bibfnamefont{S.~J.~L.} \bibnamefont{Billinge}},
  \bibinfo{author}{\bibfnamefont{A.~K.} \bibnamefont{Cheetham}},
  \bibinfo{author}{\bibfnamefont{D.}~\bibnamefont{Louca}}, \bibnamefont{and}
  \bibinfo{author}{\bibfnamefont{J.~B.} \bibnamefont{Parise}},
  \bibinfo{journal}{Appl. Phys. A} \textbf{\bibinfo{volume}{74}},
  \bibinfo{pages}{s163} (\bibinfo{year}{2002}).

\bibitem[{\citenamefont{Mathur}(1997)}]{mathu;n97}
\bibinfo{author}{\bibfnamefont{N.}~\bibnamefont{Mathur}},
  \bibinfo{journal}{Nature} \textbf{\bibinfo{volume}{390}},
  \bibinfo{pages}{229} (\bibinfo{year}{1997}).

\bibitem[{\citenamefont{Lynn et~al.}(2007)\citenamefont{Lynn, Argyriou, Ren,
  Chen, Mukovskii, and Shulyatev}}]{lynn;prb07}
\bibinfo{author}{\bibfnamefont{J.~W.} \bibnamefont{Lynn}},
  \bibinfo{author}{\bibfnamefont{D.~N.} \bibnamefont{Argyriou}},
  \bibinfo{author}{\bibfnamefont{Y.}~\bibnamefont{Ren}},
  \bibinfo{author}{\bibfnamefont{Y.}~\bibnamefont{Chen}},
  \bibinfo{author}{\bibfnamefont{Y.~M.} \bibnamefont{Mukovskii}},
  \bibnamefont{and} \bibinfo{author}{\bibfnamefont{D.~A.}
  \bibnamefont{Shulyatev}}, \bibinfo{journal}{Phys. Rev. B}
  \textbf{\bibinfo{volume}{76}}, \bibinfo{pages}{014437}
  (\bibinfo{year}{2007}).

\bibitem[{\citenamefont{Jir\'{a}k et~al.}(1985)\citenamefont{Jir\'{a}k,
  Krupi\v{c}ka, \v{S}im\v{s}a, Dlouh\'a, and Vratislav}}]{jirak;jmmm85}
\bibinfo{author}{\bibfnamefont{Z.}~\bibnamefont{Jir\'{a}k}},
  \bibinfo{author}{\bibfnamefont{S.}~\bibnamefont{Krupi\v{c}ka}},
  \bibinfo{author}{\bibfnamefont{Z.}~\bibnamefont{\v{S}im\v{s}a}},
  \bibinfo{author}{\bibfnamefont{M.}~\bibnamefont{Dlouh\'a}}, \bibnamefont{and}
  \bibinfo{author}{\bibfnamefont{S.}~\bibnamefont{Vratislav}},
  \bibinfo{journal}{J. Magn. Magn. Mater.} \textbf{\bibinfo{volume}{53}},
  \bibinfo{pages}{153} (\bibinfo{year}{1985}).

\bibitem[{\citenamefont{Elovaara et~al.}(2014)\citenamefont{Elovaara, Huhtinen,
  Majumdar, and Patur}}]{elova;jpcm14}
\bibinfo{author}{\bibfnamefont{T.}~\bibnamefont{Elovaara}},
  \bibinfo{author}{\bibfnamefont{H.}~\bibnamefont{Huhtinen}},
  \bibinfo{author}{\bibfnamefont{S.}~\bibnamefont{Majumdar}}, \bibnamefont{and}
  \bibinfo{author}{\bibfnamefont{P.}~\bibnamefont{Patur}}, \bibinfo{journal}{J.
  Phys.: Condens. Mat.} \textbf{\bibinfo{volume}{26}}, \bibinfo{pages}{266005}
  (\bibinfo{year}{2014}).

\bibitem[{\citenamefont{Wu et~al.}(2001)\citenamefont{Wu, Ogale, Nagaraj,
  Amlan, Chen, Greene, Ramesh, Venkatesan, and Millis}}]{wu;prl01}
\bibinfo{author}{\bibfnamefont{T.}~\bibnamefont{Wu}},
  \bibinfo{author}{\bibfnamefont{S.~B.} \bibnamefont{Ogale}},
  \bibinfo{author}{\bibfnamefont{B.}~\bibnamefont{Nagaraj}},
  \bibinfo{author}{\bibfnamefont{B.}~\bibnamefont{Amlan}},
  \bibinfo{author}{\bibfnamefont{Z.}~\bibnamefont{Chen}},
  \bibinfo{author}{\bibfnamefont{R.~L.} \bibnamefont{Greene}},
  \bibinfo{author}{\bibfnamefont{R.}~\bibnamefont{Ramesh}},
  \bibinfo{author}{\bibfnamefont{T.}~\bibnamefont{Venkatesan}},
  \bibnamefont{and} \bibinfo{author}{\bibfnamefont{A.~J.}
  \bibnamefont{Millis}}, \bibinfo{journal}{Phys. Rev. Lett.}
  \textbf{\bibinfo{volume}{86}}, \bibinfo{pages}{5998} (\bibinfo{year}{2001}).

\bibitem[{\citenamefont{Tao et~al.}(2005)\citenamefont{Tao, Niebieskikwiat,
  Salamon, and Zuo}}]{tao;prl05}
\bibinfo{author}{\bibfnamefont{J.}~\bibnamefont{Tao}},
  \bibinfo{author}{\bibfnamefont{D.}~\bibnamefont{Niebieskikwiat}},
  \bibinfo{author}{\bibfnamefont{M.~B.} \bibnamefont{Salamon}},
  \bibnamefont{and} \bibinfo{author}{\bibfnamefont{J.~M.} \bibnamefont{Zuo}},
  \bibinfo{journal}{Phys. Rev. Lett.} \textbf{\bibinfo{volume}{94}},
  \bibinfo{pages}{147206} (\bibinfo{year}{2005}).

\end{thebibliography}
\end{document}